\documentclass[twocolumn,superscriptaddress,floatfix,preprintnumbers, nofootinbib,hyperref]{revtex4-2} 
\pdfoutput=1
\usepackage[colorlinks=true,breaklinks=true]{hyperref}
\usepackage[normalem]{ulem}
\usepackage[utf8]{inputenc}
\hypersetup{allcolors=[rgb]{0.0 0.0 0.6},linkcolor=[rgb]{0.75 0.05 0.05}}
\usepackage{amsmath,amssymb}
\usepackage{epsfig}  
\usepackage{graphicx}   
\usepackage{slashed}       
\usepackage{tikz}
\usepackage{color}
\usepackage{tikz-feynman}
\usepackage{url}
\usepackage{array}
\newcolumntype{L}[1]{>{\raggedright\let\newline\\\arraybackslash\hspace{0pt}}m{#1}}
\newcolumntype{C}[1]{>{\centering\let\newline\\\arraybackslash\hspace{0pt}}m{#1}}
\newcolumntype{R}[1]{>{\raggedleft\let\newline\\\arraybackslash\hspace{0pt}}m{#1}}
\usepackage{multirow}
\usepackage{colortbl}
\usepackage{comment}
\usepackage{amssymb}

\DeclareMathOperator{\cm}{cm}
\DeclareMathOperator{\GeV}{GeV}

\DeclareMathOperator{\MeV}{MeV}

\DeclareMathOperator{\s}{s}

\DeclareMathOperator{\km}{km}

\DeclareMathOperator{\kpc}{kpc}

\newcommand{\beq}{\begin{equation}}
\newcommand{\eeq}{\end{equation}}

\allowdisplaybreaks

\bibpunct{[}{]}{,}{n}{}{,}

\begin{document}

\title{Multimessenger search for electrophilic feebly interacting particles from supernovae}

\author{Pedro De la Torre Luque}\email{pedro.delatorreluque@fysik.su.se}
\affiliation{The Oskar Klein Centre, Department of Physics, Stockholm University, Stockholm 106 91, Sweden}

\author{Shyam Balaji}
\email{sbalaji@lpthe.jussieu.fr}
\affiliation{Laboratoire de Physique Th\'{e}orique et Hautes Energies (LPTHE),
UMR 7589 CNRS \& Sorbonne Universit\'{e}, 4 Place Jussieu, F-75252, Paris, France}
\affiliation{Institut d'Astrophysique de Paris, UMR 7095 CNRS \& Sorbonne Universit\'{e}, 98 bis boulevard Arago, F-75014 Paris, France}

\author{Pierluca Carenza}\email{pierluca.carenza@fysik.su.se}
\affiliation{The Oskar Klein Centre, Department of Physics, Stockholm University, Stockholm 106 91, Sweden}

\smallskip

\begin{abstract}
We study MeV-scale electrophilic Feebly Interacting Particles (FIPs), that may be abundantly produced in Supernova (SN) explosions, escape the star and decay into electrons and positrons. This exotic injection of leptons in the Milky Way leaves an imprint in both photon and cosmic-ray fluxes. Specifically, positrons lose energy and annihilate almost at rest with background electrons, producing photons with $511$~keV energy. In addition, electrons and positrons radiate photons through bremsstrahlung emission and upscatter the low-energy galactic photon fields via the inverse Compton process generating a broad emission from X-ray to $\gamma$-ray energies. Finally, electrons and positrons are directly observable in cosmic ray experiments.
In order to describe the FIP-induced lepton injection in full generality, we use a model independent parametrization which can be applied to a host of FIPs such as  axion-like particles, dark photons and sterile neutrinos. Theoretical predictions are compared to experimental data to robustly constrain FIP-electron interactions with an innovative multimessenger analysis.  
\end{abstract}

\maketitle

\section{Introduction}
Explosions of massive stars, called Supernovae (SN), produce a copious amount of light Feebly Interacting Particles (FIPs). Popular and well studied examples of such new particles are axions~\cite{Raffelt:1987yt,Keil:1996ju,Carenza:2019pxu,Carenza:2020cis} and axion-like particles (ALPs)~\cite{Caputo:2021rux,Lella:2022uwi,Lella:2023bfb}, light CP-even scalars~\cite{Balaji:2022noj,Balaji:2023nbn,Dev:2021kje,Dev:2020jkh,Dev:2020eam}, sterile neutrinos~\cite{Kolb:1996pa,Raffelt:2011nc,Mastrototaro:2019vug}, dark photons (DPs)~\cite{Chang:2016ntp}, dark flavored particles~\cite{Camalich:2020wac} and unparticles~\cite{Hannestad:2007ys} (see Ref.~\cite{Antel:2023hkf} for a broad discussion on FIPs).
Therefore, SN play a fundamental role in understanding the existence and properties of FIPs. 
For typical SN core temperatures $\mathcal{O}(30)$ MeV, FIPs with masses up to $\mathcal{O}(100)$ MeV can be abundantly emitted. 
The weak interactions between FIPs and ordinary matter allow efficient subtraction of energy from the stellar core, thereby shortening the
neutrino burst signal~\cite{Raffelt:2012kt,Chang:2018rso}. This powerful physical argument has been used to exclude different new physics scenarios based on the SN 1987A neutrino observation~\cite{Raffelt:1987yt,Caputo:2022rca}. 

FIPs don't only indirectly affect the SN neutrino burst, but they also leave various signatures depending on their interactions or decays outside the SN.
For example, very light ALPs that interact with photons can be converted into $\gamma$-rays in the galactic magnetic field. This would lead to a distinctive $\gamma$-ray emission in coincidence with the SN explosion. This approach has been used to search for ALPs from SN 1987A~\cite{Payez:2014xsa,Hoof:2022xbe} and to predict their signals from future galactic or extra-galactic SN~\cite{Meyer:2016wrm,Meyer:2020vzy,Crnogorcevic:2021wyj,Calore:2023srn}, as well as from the diffuse $\gamma$-ray background \cite{Calore:2020tjw,Calore:2021hhn,Balaji:2022wqn}.
A similar idea can be applied to MeV-scale ALPs, decaying into photons after escaping the SN and generating an unexpected $\gamma$-ray emission~\cite{Giannotti:2010ty,Jaeckel:2017tud,Caputo:2021kcv,Hoof:2022xbe,Muller:2023vjm,Muller:2023pip,Diamond:2023scc}.

In this work we focus on electrophilic FIPs with masses in the MeV-range, particularly interesting for phenomenology~\cite{Appelquist:2002me,Asaka:2005an,Abel:2008ai,Jaeckel:2010ni} and experimental searches~\cite{Alekhin:2015byh}.
We explore the multimessenger signals induced by FIPs decaying into electron-positron pairs, showing that they can be used to constrain a variety of well-motivated FIP models, such as ALPs, DPs and sterile neutrinos.
FIPs produced in a SN and decaying into electrons and positrons, contribute to the total galactic background of electrons and positrons generated from astrophysical sources (mainly pulsar wind nebulas and SN remnants)~\cite{1969ocr..book.....G,Gabici:2019jvz} and cosmic ray (CR) interactions with the interstellar gas~\cite{DelaTorreLuque:2023zyd}.

An observable strongly affected by this exotic lepton injection is the $511$~keV signal generated from positron interactions with interstellar gas. 
We exploit observations at these energies to set limits on the total injected flux of electrons and positrons, agnostic to the exact FIP model. The proposed recipe can be readily applied to any electrophilic FIP efficiently produced in SN.
In the pioneering work Ref.~\cite{Dar:1986wb} (also see Sec.~12.5.1 of Ref.~\cite{Raffelt:1996wa}), the authors examined the $511$~keV photon constraint for decaying heavy neutrinos, using early data on the $511$~keV observations. The same bound for DPs was recently explored in Ref.~\cite{DeRocco:2019njg}. The best constraints to date on different FIP models come from this observable~\cite{Calore:2021klc,Calore:2021lih}. We improve upon these previous studies by using  updated data from the  Spectrometer on INTEGRAL (SPI) instrument~\cite{Bouchet:2010dj,Siegert:2015knp} and exploring a wider set of models for the spatial distribution of SN, revising the systematic uncertainties in these evaluations.

In this paper, extending the companion paper Ref.~\cite{DelaTorreLuque:2023nhh}, we propose alternative constraints on the FIP-induced electron-positron injection. For instance, in recent years the Voyager-1 probe has provided measurements of the local flux of electrons outside the heliosphere~\cite{Cummings:2016pdr,2013Sci...341..150S} at energies below tens of MeV. These measurements profer the advantage that they are not significantly affected by the solar modulation effect~\cite{Potgieter:2013pdj}, which highly suppresses the flux of low-energy charged CRs at Earth.
For the first time, we use the Voyager-1 electron data to constrain FIPs.
Likewise, the secondary emissions from the electron-positron pairs, especially from Inverse Compton (IC) and bremsstrahlung processes, have been studied in the context of light dark matter recently~\cite{Cirelli:2020bpc} but has not yet been applied to the case of FIPs. We explore these secondary emission signals and perform a systematic analysis of the available datasets for hard X-rays and low-energy $\gamma$ rays, to determine their full constraining power on electrophilic FIPs.
The various data explored in our multimessenger approach allow us to set stringent and robust bounds on electrophilic FIPs, competitive with limits from $511$~keV observations. 

This manuscript is organized as follows. In Section~\ref{sec:FIPphen} we introduce the phenomenology of electrophilic FIPs and characterize their production spectrum from SN. In Section~\ref{sec:El_flux_FIPs}, we give details on the FIP-induced injection and diffusion of electrons and positrons in the Galaxy. Then, their secondary photon emissions are briefly discussed in Section~\ref{sec:511calculation}.
The various datasets, at different energies, used to constrain FIP properties are discussed in Section~\ref{sec:constraints}. To summarize, we discuss constraints from:
\begin{itemize}
    \item the $511$~keV line measured by the SPectrometer on INTEGRAL (SPI, Sec.~\ref{sec:511});
    \item local electron and positron fluxes detected by Voyager-1 and Alpha Magnetic Spectrometer (AMS-02, Sec.~\ref{sec:voyAMS});
    \item high-energy $\gamma$-rays in the $20$~MeV-$30$~GeV range probed by the Energetic Gamma Ray Experiment Telescope (EGRET);
    \item $\gamma$-rays with energies $2$-$20$~MeV revealed by the Imaging Compton Telescope (COMPTEL);
    \item soft $\gamma$-rays and hard X-rays, between $20$~keV and $2$~MeV, covered by SPI;
    \item X-rays in the $2.5-8$~keV range, studied with the X-ray Multi-Mirror Mission (XMM-Newton);
\end{itemize}
where the X- to $\gamma$-ray signals are discussed in Section~\ref{sec:XGamma}. These constraints are applied to various FIP models in Section~\ref{sec:FIPconstr}. In Section~\ref{sec:conclusions} we summarize the main results and conclude.

\section{Electrophilic FIP phenomenology}
\label{sec:FIPphen}

A large class of FIPs, dubbed electrophilic, feature a coupling to electrons and positrons. An electrophilic FIP, $X$, with a mass larger than the electron-positron energy threshold is kinematically able to produce an electron-positron pair in its decay. For example, ALPs ($a$) with $m_{a}\gtrsim 1$~MeV decay as $a\to e^{+}e^{-}$ and similarly for DPs. Also massive sterile neutrinos produce electrons and positrons in their decays, for example channels such as $\nu_{s}\to \nu_{\mu}e^{+}e^{-}$ are open for sterile neutrinos mixed with muon neutrinos, and other decays are possible as the sterile neutrino mass is increased. Several other FIPs can also produce electrons and positrons in their decay, like scalars, Kaluza-Klein gravitons and supersymmetric particles. Thus, it is important to search for signatures of electrophilic FIPs.

Once produced in a SN, FIPs may escape from the progenitor star subtracting energy from the core due to their weak interactions with the stellar medium. The energy lost from the star might be comparable with losses accounted for by neutrinos which significantly affect the SN evolution. Hence, based on the observation of SN 1987A, it is possible to extract stringent bounds that apply to several FIP models since cooling due to FIPs should not significantly affect the SN dynamics. This energy-loss criterion can be cast as a constraint on the FIP luminosity~\cite{Raffelt:1990yz}
\begin{equation}
    L_{X}^{\rm tot}\lesssim3\times10^{52}\,{\rm erg}~{\rm s}^{-1}\,,
\end{equation}
at the beginning of the cooling phase, at about $t_{\rm pb}=1$~s, where $t_{\rm pb}$ is the post-bounce time.
\begin{figure*}[t!]
\begin{minipage}{.47\textwidth}
    \includegraphics[width=\columnwidth]{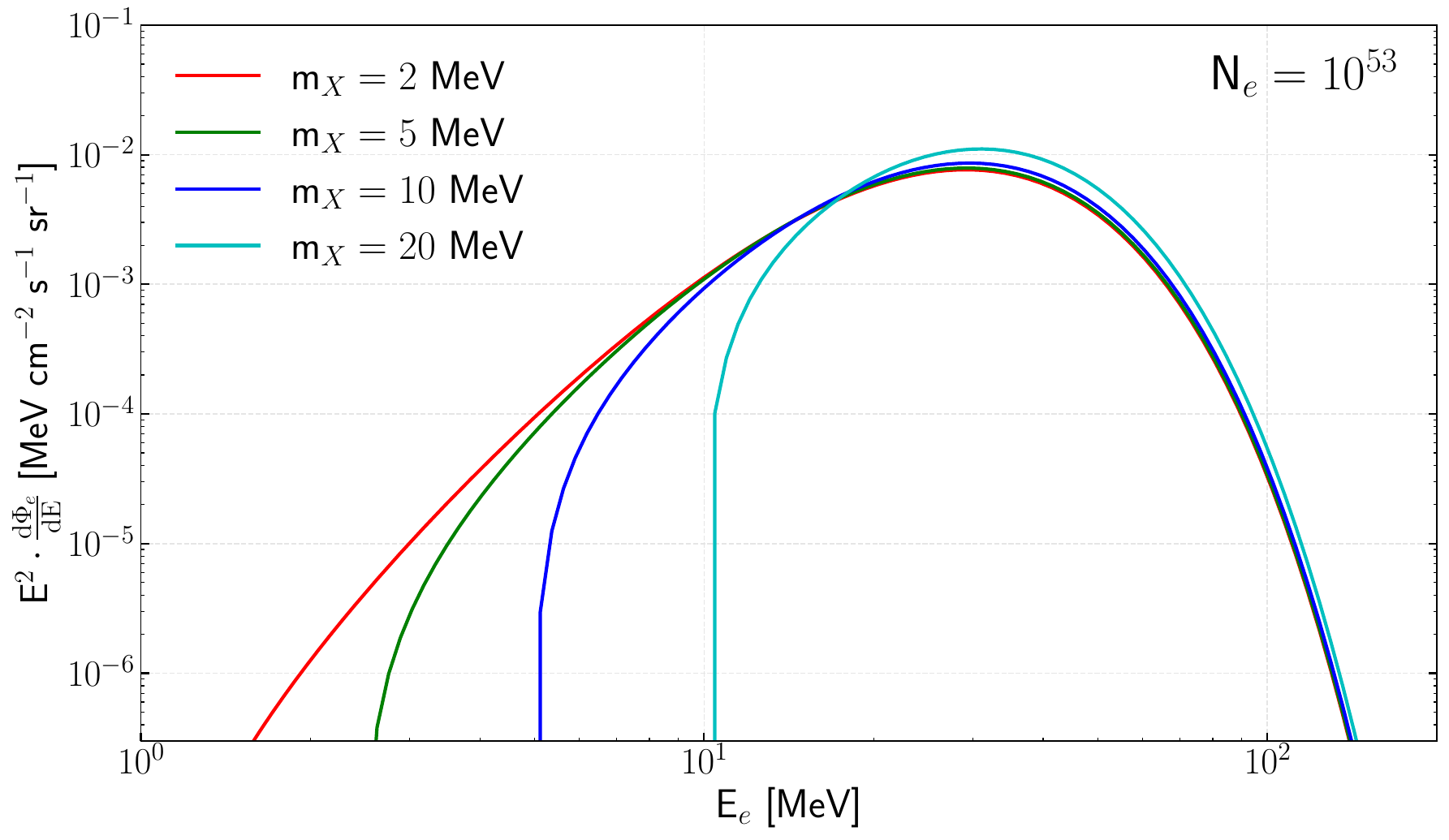}
  \end{minipage} \quad
  \hspace{0.1cm}
  \begin{minipage}{.47\textwidth}
    \includegraphics[width=\columnwidth]{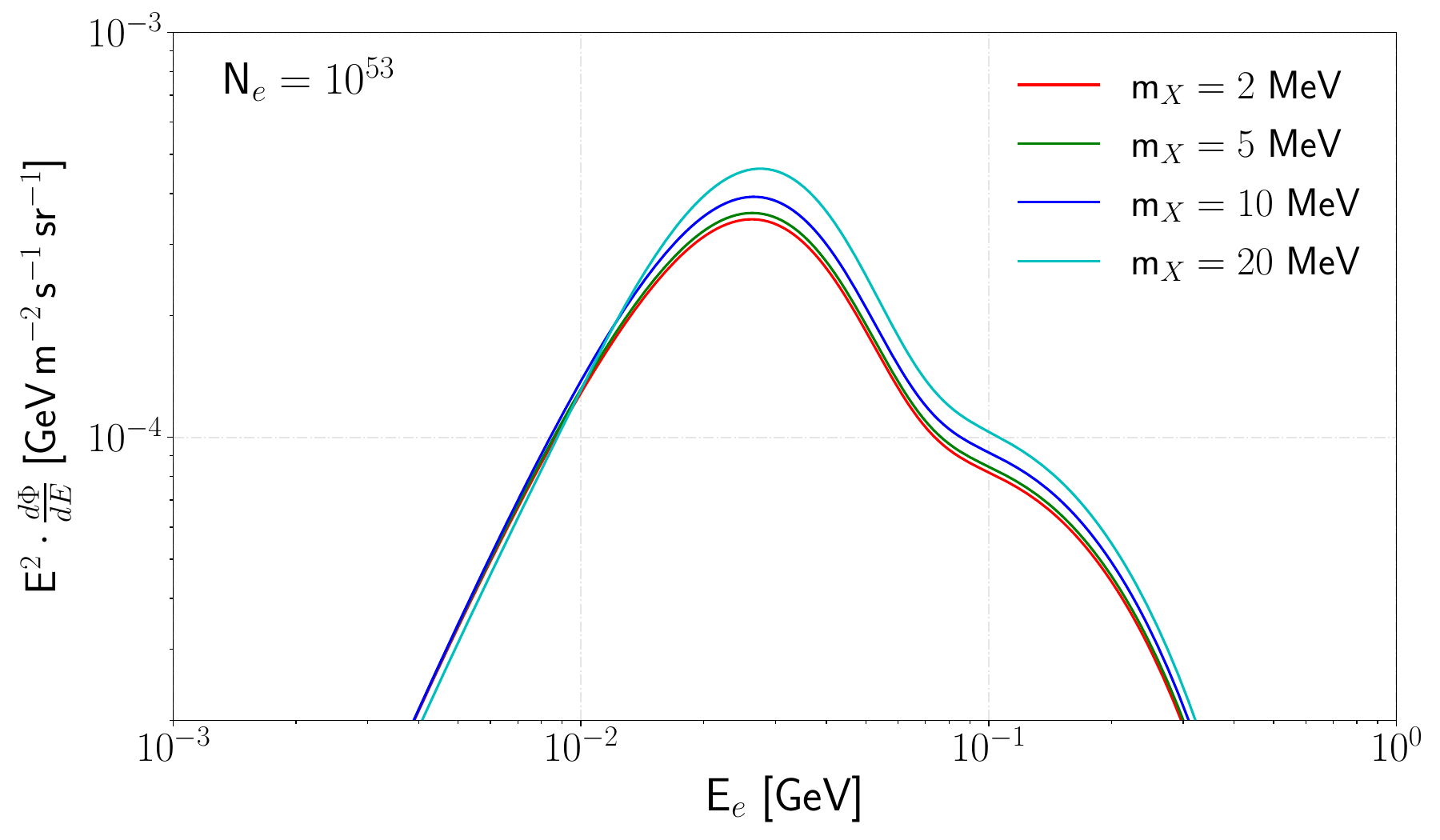}
  \end{minipage} \quad
\caption{{\it Left panel}: Total injected flux of electrons for different FIP masses and $N_e=10^{53}$, shown as function of the total energy of the electron. {\it Right panel}: Local (unmodulated) flux of the propagated electron population obtained by injecting the FIP-induced fluxes in the left panel.}
\label{fig:Inj}
\end{figure*}
In the case that FIPs decay electromagnetically, a more severe constraint was obtained in Ref.~\cite{Caputo:2022mah} (see also Refs.~\cite{Falk:1978kf,Sung:2019xie}) from the observation of low-energy SN, requiring that the energy injected by FIP decay products satisfies
\begin{equation}
    L_{X}^{\rm e.m.}\lesssim10^{50}\,{\rm erg}~{\rm s}^{-1}\,.
\end{equation}
Here, with the superscript e.m., we specified that this constraint applies to FIP luminosity converted into an electromagnetic channel.

This bound is valid for FIPs decaying inside the SN envelope, FIPs with longer mean-free-paths cannot be constrained. However, in this case, the copious FIP production in a SN is expected to inject a large flux of $\gamma$-rays, electrons and positrons in our Galaxy. Typically the electron/positron flux will inherit the spectral properties of the parent particle. If the decay products have an energy $E_{e}$ which is half of the parent FIP, the injected electron/positron flux can be written  as~\cite{vanBibber:1988ge}
\begin{equation}
\begin{split}
    \frac{dN_{e}}{dE_{e}}&=N_{e}C_{0}\left(\frac{4E_{e}^{2}-m_{X}^{2}}{E_{0}^{2}}\right)^{\beta/2}e^{-(1+\beta)\frac{2E_{e}}{E_{0}}}\,,\\
      C_{0}&=\frac{2\sqrt{\pi}\left(\frac{1+\beta}{2m_{X}}\right)^{\frac{1+\beta}{2}}E_{0}^{\frac{\beta-1}{2}}}{K_{\frac{1+\beta}{2}}\left((1+\beta)\frac{m_{X}}{E_{0}}\right)\Gamma\left(1+\frac{\beta}{2}\right)}\,,
\end{split}
\label{eq:spectrum}
\end{equation}
assuming that FIPs are emitted with a modified blackbody spectrum, where $E_{0}$ is related to the FIP average energy, its mass is $m_{X}>2m_{e}$, $\beta$ is the spectral index, $K_{\frac{1+\beta}{2}}$ is the modified Bessel function of the second kind of order $(1+\beta)/2$, $\Gamma$ is the Euler-Gamma function and this flux is normalized such that 
\begin{equation}
    \int_{m_{X}/2}^{\infty} dE_{e}\frac{dN_{e}}{dE_{e}}=N_{e}\,.
\end{equation}
Throughout this paper, we use $N_{e}$ to denote the number of electrons, which is equal to the number of positrons, produced in a SN explosion via FIP decays, i.e.~$N_{e}=N_{e^{+}}=N_{e^{-}}$. 
The simple prescription in Eq.~\eqref{eq:spectrum} does not depend on the type of FIP model.
This flux is obtained by assuming a FIP decaying into an electron-positron pair. Thus, it cannot be strictly valid for sterile neutrinos, which has more involved decay channels.
However, by changing the parameters we expect that sterile neutrino spectra can also be roughly modeled by Eq.~\eqref{eq:spectrum}. 
This simple prescription for the injected lepton flux enables us to place constraints on the electron-coupling of different electrophilic FIPs in a model-independent way. Unless otherwise specified, in this work we adopt the values of $E_0 =45$~MeV, $\beta = 2.5$ and $m_{X}=10$~MeV as our benchmark model.

We will now discuss in detail the non-standard injection of electrons and positrons in the Galaxy, caused by FIP decays, and its relevant phenomenological consequences.

\begin{figure*}[t!]
 \begin{minipage}{.5\textwidth}
    \includegraphics[width=\linewidth]{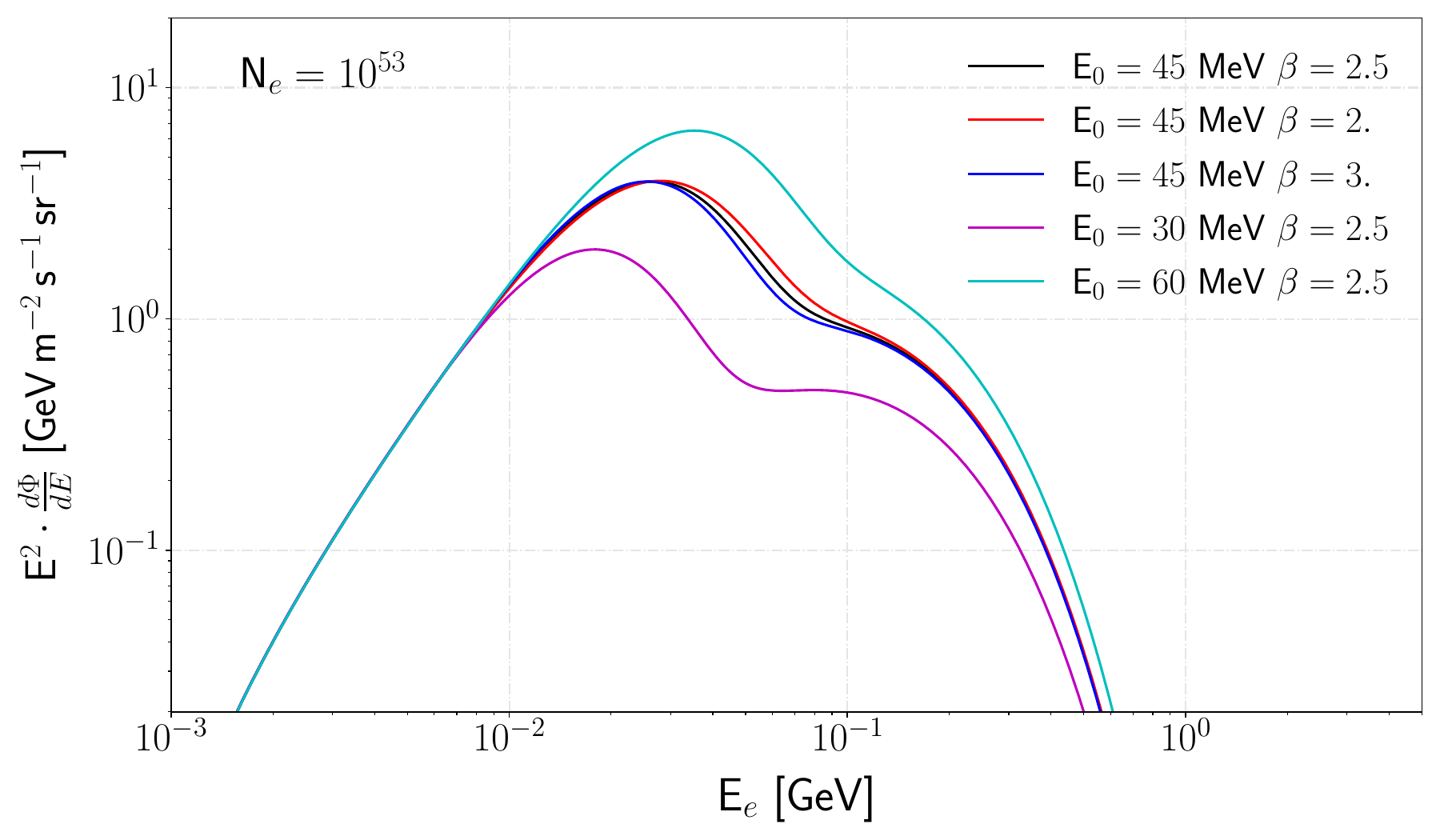}
  \end{minipage}
   \hspace{0.1cm}
  \begin{minipage}{.38\textwidth}
    \includegraphics[width=\linewidth]{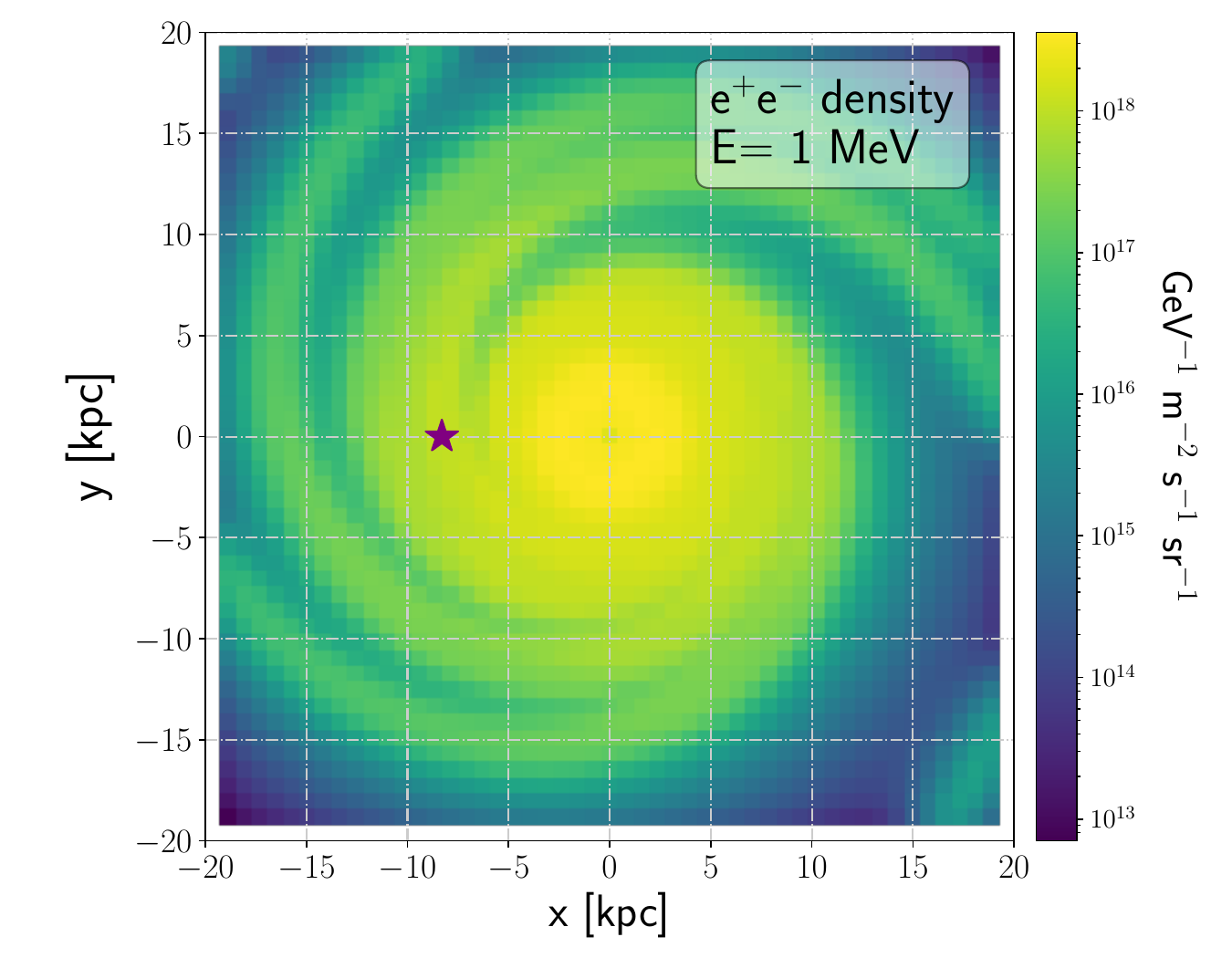}
  \end{minipage} \quad
 
  \quad
  \caption{{\it Left panel}: Electron spectrum at the Earth location for different injection parameters, see Eq.~\eqref{eq:spectrum}. In both panels we use $N_e=10^{53}$ and a FIP mass $m_{X}=10$~MeV. {\it Right panel}: Galactic map representing the density of propagated electrons generated from FIPs, at an energy of $1$~MeV, and in galactic coordinates. We have added a star marker to indicate the Solar System position for clarity.}
  \label{fig:InjEl}
\end{figure*}

\subsection{Galactic electron-positron flux induced by electrophilic FIPs}
\label{sec:El_flux_FIPs}

In order to calculate the spectrum and galactic distribution of electrons and positrons produced in FIP decays, we solve the diffusion equation of these particles using the {\tt DRAGON2} code~\cite{Evoli:2016xgn,Evoli:2017vim}.\footnote{\label{note1}\url{https://github.com/cosmicrays/DRAGON2-Beta\_version}} {\tt DRAGON2} is an advanced CR propagation code designed to self-consistently solve the diffusion-advection-loss equation describing CR transport for all species involved in the CR network, including CRs of both astrophysical and exotic origin (e.g.,  from dark matter annihilations/decays). 
\begin{table}[t!]
    \centering
    \begin{tabular}{|c|c|c|}
    \hline
Norm. Energy    &    $E_{0}$ & $4~\GeV$ \\
Diffusion coeff.&    $D_{0}$  & $1.02\times10^{29}\cm^{2}\s^{-1}$ \\
Diffusion index &    $\delta$&$0.49$  \\
Break energy &       $E_{b}$&$312~\GeV$ \\
Index break &        $\Delta\delta$&$0.20$ \\
Smooth. param. &     $s$&$0.04$\\
$\beta$ exponent &  $\eta$&      $-0.75$\\
Halo height&      $H$&$8~\kpc$\\
Alfvèn velocity&  $v_{A}$&$13.4\km\s^{-1}$\\
\hline
\end{tabular}
\caption{Main propagation parameters used in our analysis~\cite{DelaTorreLuque:2023zyd}. }
\label{tab:params}
\end{table}
We work under the hypothesis that FIPs decay relatively close, at a distance smaller than $\sim\mathcal{O}(10)~\kpc$~\cite{Calore:2021klc}, to the SN where they are produced. Thus, we simulate the injection of electrons and positrons as happening directly from SN distributed following the spatial distribution determined by the Ferriere distribution~\cite{Ferriere:2007yq}, `Ferr', convolved with the Steiman-Cameron distribution~\cite{Steiman-Cameron:2010iuq} of the spiral arms (four-arm model), and the injection spectrum given by Eq.~\eqref{eq:spectrum}. Alternatively, we use the Lorimer distribution~\cite{Lorimer:2006qs}, `Lor', in order to estimate the uncertainties related to the SN distribution.
The electron/positron injection flux is an input of the code. We define it as 
\begin{equation}
\frac{d\Phi_e}{dE_e}=\frac{\Gamma_{cc}}{4\pi d_{SN}^2}  \frac{dN_{e}}{dE_{e}} \, ,
    \label{eq:injection}
\end{equation}
where $\Gamma_{cc} = 2$ SNe per century is the rate of core-collapse SN explosions in the Galaxy~\cite{Rozwadowska:2020nab}\footnote{Note that the reported value of the Galactic SN explosion rate is $\Gamma_{cc} =  1.63 \pm 0.46$ SNe per century~\cite{Rozwadowska:2020nab}. Thus, the assumed flux might be halved in a pessimistic case corresponding to $\Gamma_{cc}\sim1$ SN per century. 
Since the number of positrons per SN is proportional to the FIP-electron coupling squared, in the low-coupling regime, this would reflect in a reduction of a factor $\sim \sqrt{2}$ on the constrained FIP-electron coupling. The strong coupling regime is even less sensitive to this relaxation because of the steep dependence of $N_{e}$ on the FIP-electron coupling. Therefore, this uncertainty is smaller than others, associated with Galactic propagation, in this type of study (see discussion in Ref.~\cite{DelaTorreLuque:2023nhh}).} and $d_{SN}=10.2$~kpc is an effective length-scale necessary to calculate the total area of emission from the SN. It is roughly the average distance between galactic SN calculated from the source distribution employed in this work. This parameter is obtained by imposing that the injected number of particles per unit of energy in Eq.~\eqref{eq:spectrum} is equal to the integral of the total flux density of particles (obtained convolving Eq.~\eqref{eq:spectrum} with the spatial distribution of sources) over the volume of the Galaxy.
Furthermore, we remark that electrons and positrons propagate and interact in the Galaxy on a timescale of $10^{3}$-$10^{6}$~years, extremely long compared to the SN explosion rate. Therefore, we can model the lepton injection as time-independent and smoothly following the SN distribution. 

In this setup, we employ the spatially-constant diffusion coefficient derived from the analyses of secondary-to-primary flux ratios in Ref.~\cite{delaTorreLuque:2022vhm} and adapted in Ref.~\cite{DelaTorreLuque:2023zyd} for the 3D source, gas and magnetic field distribution of the Galaxy (B/C best-fit model). In particular, for the diffusion coefficient defined in Eq.~(3.3) of Ref.~\cite{delaTorreLuque:2022vhm}, the parameters are shown in Tab.~\ref{tab:params}.
We adopt the magnetic field model derived by Ref.~\cite{Pshirkov:2011um}, with a normalization of the disk, halo and turbulent magnetic field intensities set to the values found in Ref.~\cite{DiBernardo:2012zu} from the study of synchrotron radiation. The energy density distribution of the radiation fields has been taken from Ref.~\cite{Porter:2008ve}.

In the left panel of Fig.~\ref{fig:Inj} we show the FIP-induced electron/positron flux injected for a single SN, assuming a fixed total number of injected positrons $N_{e}=10^{53}$, obtained by Eq.~\eqref{eq:injection} for various FIP masses.
These injections give rise to a galactic electron flux that, after propagation, is shown in the right panel of Fig.~\ref{fig:Inj}. Here, the flux is evaluated at the Earth location and we neglect the solar modulation effect, which is negligible outside the heliosphere.
From this comparison we note that different masses in the $2$-$20$~MeV range lead to very similar spectra at their peak. Moreover, their propagated spectra are very similar below $\sim20$~MeV, as we see from the right panel of Fig.~\ref{fig:Inj}.
Another peculiar feature is that the propagated electron/positron flux peaks at an energy slightly below the peak one of the injected flux, because of the energy lost by leptons during their propagation. Moreover, there is another peak at higher energies, around $\sim100$~MeV, due to reacceleration of electrons/positrons by plasma waves~\cite{1995ApJ...441..209H,1998ApJ...509..212S}.
We limit this analysis to $m_{X}\lesssim20$~MeV because in this range our conclusions are quite mass-independent and above this threshold the FIP production starts to be Boltzmann suppressed.
The discussion until now is only focused on the mass dependence of the FIP-induced galactic lepton flux. However, this flux depends also on the FIP spectral properties, i.e. the choice of $E_{0}$ and $\beta$ in Eq.~\eqref{eq:spectrum}. This is precisely what is shown in the left panel of Fig.~\ref{fig:InjEl}. Here, we illustrate how the choice of parameters affect the predicted electron/positron flux at a fixed FIP mass $m_{X}=10$~MeV and total number of injected positrons $N_{e}=10^{53}$, for values of $30~\MeV\le E_{0}\le60~\MeV$ and $2\le\beta\le 3$. 
The black, red and blue lines share the same $E_{0}$ and different spectral indices, pinching the spectrum as $\beta$ increases. The cyan and purple lines embrace the variability of $E_{0}$ in the interval $30$-$60$~MeV, showing a suppression of the flux for lower injection energies. 
From this comparison we conclude that the results of our analysis are not strongly affected by the precise shape of the spectrum in Eq.~\eqref{eq:spectrum}, for parameters varying in the discussed range. This makes it possible to apply our results to a large number of FIP models, without expecting major quantitative differences. For a detailed study of these constraints in realistic FIP models we defer to an upcoming publication~\cite{Balajinew}.
The spatial distribution of the lepton population is also relevant for the upcoming discussion. As expected for charged particles, this population follows the spiral structure of the Galaxy, as shown in the right panel of Fig.~\ref{fig:InjEl} for an electron energy of $1$~MeV.

\subsection{Secondary emissions from the FIP-induced electron-positron population}
\label{sec:511calculation}

During their propagation in the Galaxy, electrons/positrons injected by FIPs interact with the interstellar gas and radiation fields (ISRFs), i.~e. mainly optical and ultraviolet (UV) light from stars, infrared (IR) light from dust and the Cosmic Microwave Background (CMB) as well as the galactic magnetic field.
Together, the IC emission, from the boost of the low energy ISRF photons, and the bremsstrahlung emission, from the interaction of electrons/positrons with the interstellar gas, produce a continuous emission from X-rays at the keV scale to $\gamma$ rays at the MeV scale.
Therefore, one can benefit from MeV $\gamma$-ray observations as well as the rich data in X-ray observations in order to probe the injected lepton population. 
\begin{figure}[t!]
\includegraphics[width=0.46\textwidth]{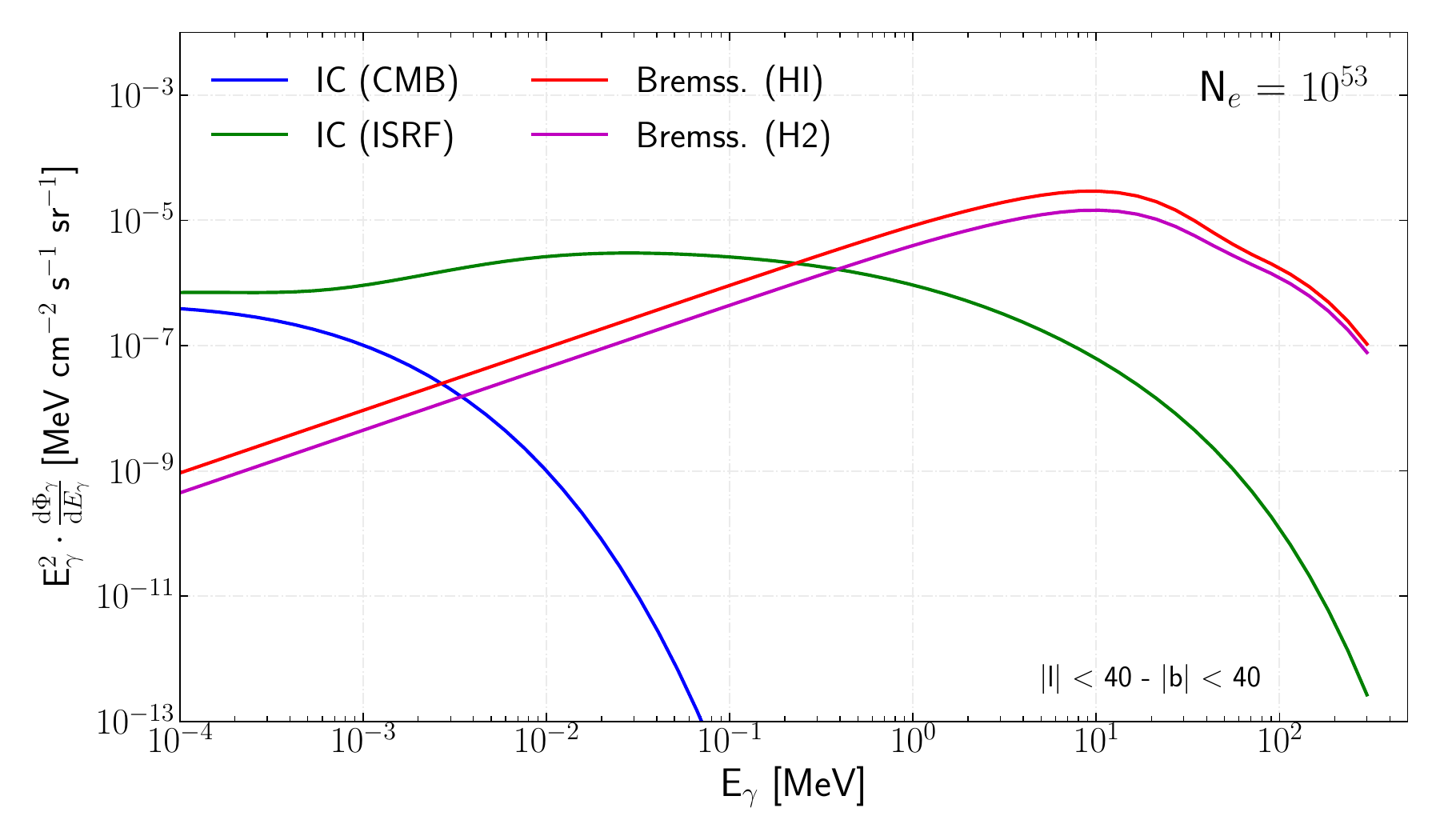}
\caption{X-to-$\gamma$-ray emission from a FIP with mass $m_{X}=10$~MeV in a region around the galactic center ($|l|<40^{\circ}$ and $ |b| < 40^{\circ}$). Here, we show the bremsstrahlung emission from the interaction with atomic (red line, labeled as HI) and molecular (magenta line, labeled as H2) gas as well as the IC emission from the interaction with the CMB (blue line, labeled as CMB) and all the ISRFs, including CMB, IR, optical and UV fields (green line, labeled as ISRF).}
\label{fig:ICBremss}
\end{figure}
\begin{figure*}[t!]
\includegraphics[width=1\textwidth]{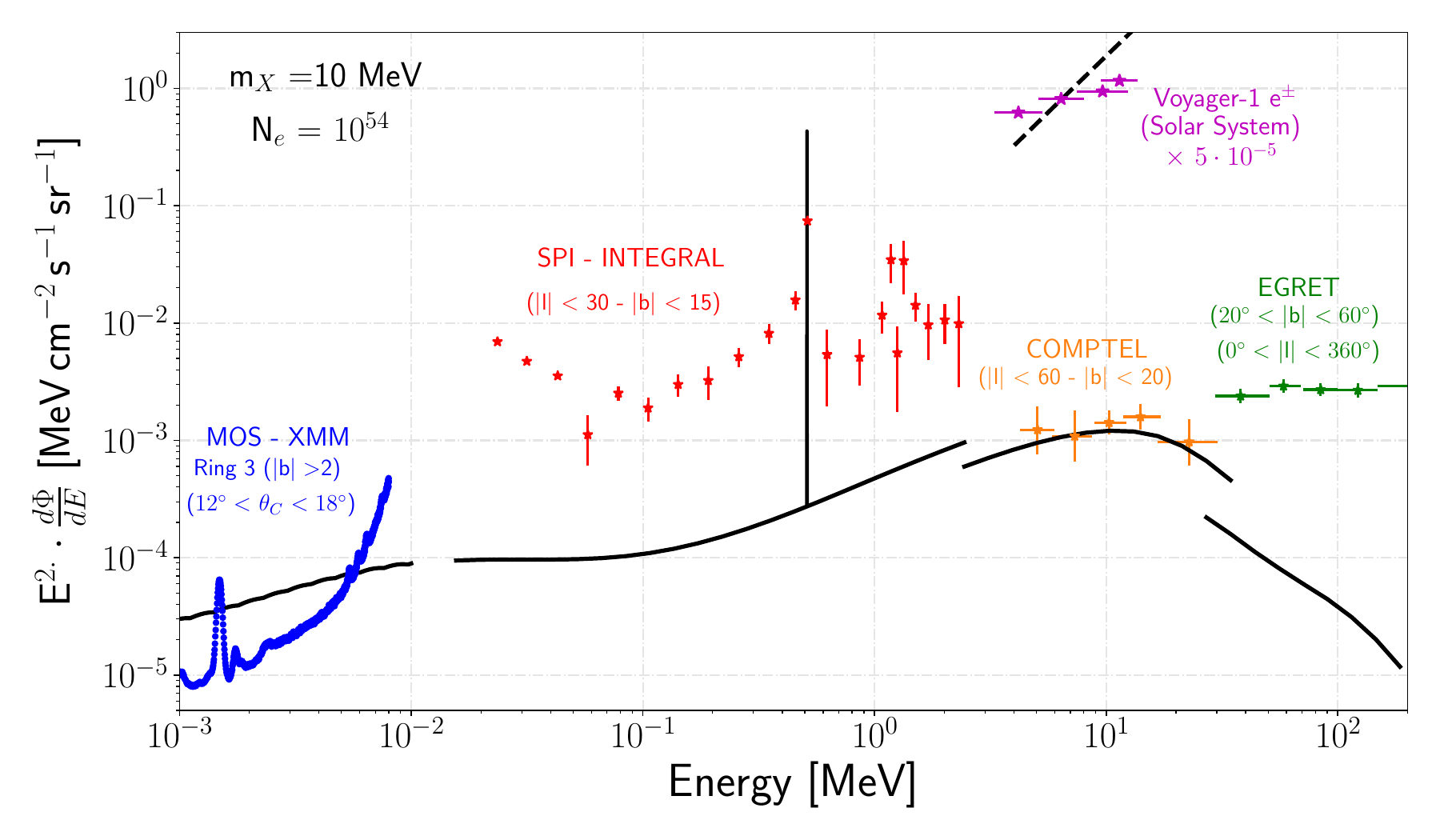}
\caption{Comparison of the predicted FIP signals, for $m_{X}=10$~MeV and $N_{e}=10^{54}$, (black lines) with the different datasets employed in this work. Since each dataset is extracted from a different region of the sky, the theoretical predictions are calculated for the spatial region and energy range of the corresponding dataset. The electron flux is indicated with a dashed line and the secondary X-to-$\gamma$-ray emission with solid lines. In addition, the calculated emission corresponding to the $511$~keV line is also shown.}
\label{fig:Sum}
\end{figure*}
In this work, we model the secondary diffuse emission due to the interactions of this electron/positron population using the {\tt HERMES} code~\cite{Dundovic:2021ryb}, which  performs a numerical integration of the emission along the line of sight at each galactic position and energy using detailed gas maps and updated ISRF models~\cite{Vernetto:2021tgp,Vernetto:2016alq}. In Fig.~\ref{fig:ICBremss} we show the spectra of the different components producing the IC and bremsstrahlung emissions from the electrons injected by a $10$~MeV FIP for a region with $|b|<40^{\circ}$ and $|l|<40^{\circ}$ in latitude and longitude around the galactic center. In particular, the IC emission from the interaction with the CMB and the rest of ISRFs is shown in blue and green, respectively, while the bremsstrahlung emission produced from the interaction with atomic and molecular gas are shown as red and magenta lines respectively. Bremsstrahlung emission, for a single scattering event, produces a flat photon spectrum that drops at a cutoff energy of the same order of magnitude of the charged particle energy. This explains the behavior of the red and magenta lines in Fig.~\ref{fig:ICBremss}. The IC emission promotes the energy of the ambient photons in the ISRFs proportionally to $E^2$. For example, a typical ISRF photon has an energy around a few eV, and it gains an energy of $(E/m_e)^2 \sim 3600$ times higher when interacting with an electron of $E\sim30$~MeV, thereby reaching a final energy at the keV-scale. At lower energies, scattering on the CMB dominates. These processes explain why the IC emission is more important at low energies while the bremsstrahlung emission peaks above a few tens of MeV.

A distinctly different process for secondary emission is the positron annihilation with galactic electrons, contributing to the $511$~keV photon line, which is a subject of notable interest~\cite{Strong:2005zx,Bouchet:2010dj,Siegert:2015knp,Siegert:2019tus}.
The FIP-injected positrons are slowed down through elastic scatterings with the interstellar gas eventually forming a positronium bound state with the ambient electrons, then annihilating almost at rest~\cite{Guessoum:2005cb}. Only around $25\%$ of positrons form (singlet) para-positronium states, which decay into two photons, each with energy almost exactly equal to the electron mass, contributing to the $511$~keV line. The (triplet) orthopositronium state, formed in $75\%$ of cases, decays into three photons with a continuous spectrum. This latter contribution will be neglected. The impact of ALPs and, subsequently of DPs and sterile neutrinos, on the $511$~keV line was discussed in Refs.~\cite{Calore:2021klc,Calore:2021lih}. 

For a given number of injected positrons $N_{e}$  it is possible to calculate the $511$~keV photon flux as
\begin{equation}
    \frac{d\phi_{\gamma}^{511}}{d\Omega}=2k_{ps}N_{\rm e}\Gamma_{cc}\int ds\,s^{2}\frac{n_{cc}(x_{s,b,l},y_{s,b,l},z_{s,b,l})}{4\pi s^{2}}\,,
\end{equation}
where $k_{ps}=1/4$ is the fraction of positronium decays contributing to the  $511$~keV line signal, $\Gamma_{cc}$ is the galactic SN rate introduced in Eq.~\eqref{eq:injection}, $n_{cc}$ is the SN density distribution, $d\Omega=dl db \cos b$ is the solid angle element with $l$ longitude and $b$ latitude and the coordinates of points at distance $s$ from the Earth contributing to the signal are determined in spherical coordinates by 
\begin{equation}
    \begin{split}
        x_{s,b,l}&=x_{0}+s \cos l\cos b\,,\\
        y_{s,b,l}&=y_{0}+s \sin l\cos b\,,\\
        z_{s,b,l}&=z_{0}+s \sin b\,,\\
    \end{split}
\end{equation}
given that the Earth is located at $(x_{0},y_{0},z_{0})=(-8.2,0,0)$~kpc~\cite{Evans:2018bqy}.
The recipes discussed in this Section to calculate secondary emissions from the injected lepton fluxes, are applied throughout this work to set constraints on $N_{e}$.

\section{Constraints on electron/positron injection by FIPs}
\label{sec:constraints}

Before discussing the specific details for the analyses of each of the datasets employed in this work, we provide a general view of the different observations and expected FIP-induced signals over a large energy range (see Fig.~\ref{fig:Sum}). In particular, the black lines are the theoretical predictions for the multimessenger signals induced by an exotic electron/positron injection of $N_{e}=10^{54}$ positrons per SN explosion. The benchmark values for the energy spectrum of the FIP decay products are assumed and $m_{X}=10$~MeV.
Our theoretical predictions are compared with various datasets, shown with different colors. Note that the black dashed line indicates the  FIP-induced electron signal (scaled by a factor $5\times10^{-5}$ for sake of clarity) that can be probed by measurements of the unmodulated electron flux by Voyager-1.
The black solid lines indicate the predicted X-to-$\gamma$-ray signal, covering the energy range from $1$~keV to $200$~MeV, including the  $511$~keV line. 
The various measurements considered are obtained in different observation regions. Thus, we compare data and FIP signals calculated in the same portion of the sky as indicated in Fig.~\ref{fig:Sum} for each dataset. In particular, we show data from the MOS detector (XMM-Newton, blue points)~\cite{Foster:2021ngm}, SPI detector on INTEGRAL (red points)~\cite{Berteaud_2022, Bouchet:2008rp}, COMPTEL (orange points)~\cite{1998PhDT.........3K} and EGRET telescope (green points)~\cite{Strong:2004de}, representing the X-to-$\gamma$-ray emission and Voyager-1 (purple points)~\cite{2013Sci...341..150S}, representing the direct electron signal.
Fig.~\ref{fig:Sum} helps to illustrate the behavior of the FIP-induced signals as a function of energy and how they compare with data. For instance, the $\gamma$-ray signal above $\sim20$~MeV is extremely low compared to EGRET observations because of the strong suppression of the bremsstrahlung emission at high energies. Therefore, we do not expect to place competitive constraints with this instrument.
At lower energies, like in the $2$-$20$~MeV range, the bremsstrahlung emission produces a sizable photon flux, peaked at $\sim 10$~MeV and iscomparable with the COMPTEL measurements.
As the energy is lowered, between $20$~keV and $2$~MeV, the IC on ISRFs becomes responsible for a large fraction of the X-ray signal. Here, SPI data for the broad emission do not constitute a strong constraint. By contrast, SPI measurements of the $511$~keV line set the leading constraints. 
Finally, in the $1$-$10$~keV region where MOS data are available, the X-ray emission is dominated by IC emission on CMB and IR photon background. These low-energy observations are very constraining, as evident from Fig.~\ref{fig:Sum}. 
For comparison, in Fig.~\ref{fig:Sum} we also show the primary electron flux at MeV energies and the Voyager-1 measurements, competitive to X-ray searches in probing exotic electron injections. 

We consider each data set and obtain bounds through a $\chi^2$ fit, imposing the $2\sigma$ bound on the parameter space whenever we obtain $\sum_i \left( \frac{\textrm{Max}\left[\phi_{\textrm{X} i}(m_{X}) - \phi_{i}, 0 \right]}{\sigma_i}\right)^2 =4$ , where $i$ denotes the data point, $\phi_{i}$ is the observed flux and $\sigma_i$ the associated standard deviation of the measurements.

\vspace{0.3cm}
In the following, we describe our analysis and results for every energy range and dataset considered.

\subsection{$511$~keV signal}
\label{sec:511}

\begin{figure*}[t!]
   \begin{minipage}{.45\textwidth}
    \includegraphics[width=\linewidth]{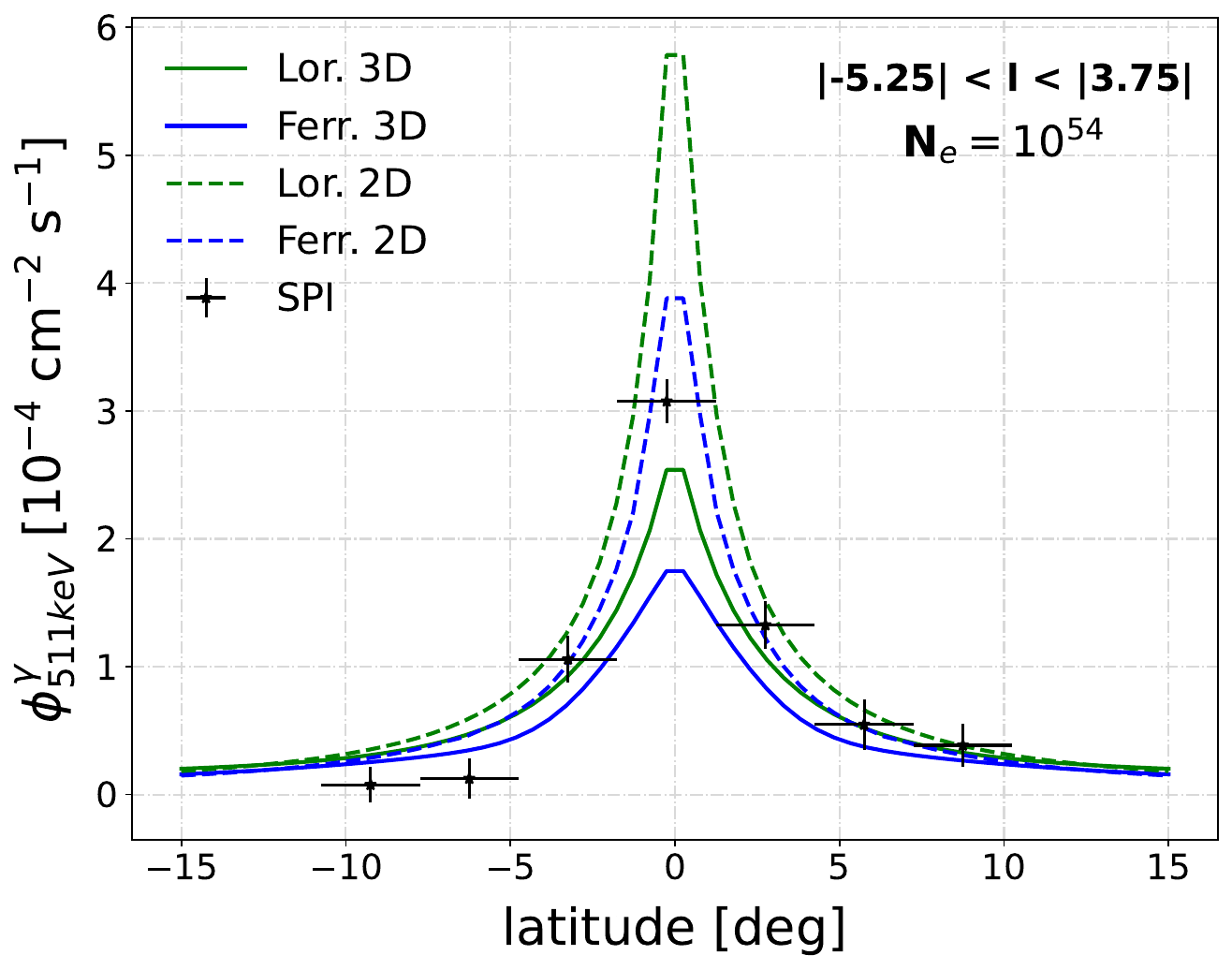}
  \end{minipage} \quad
  \hspace{0.1cm}
  \begin{minipage}{.45\textwidth}
    \includegraphics[width=\linewidth]{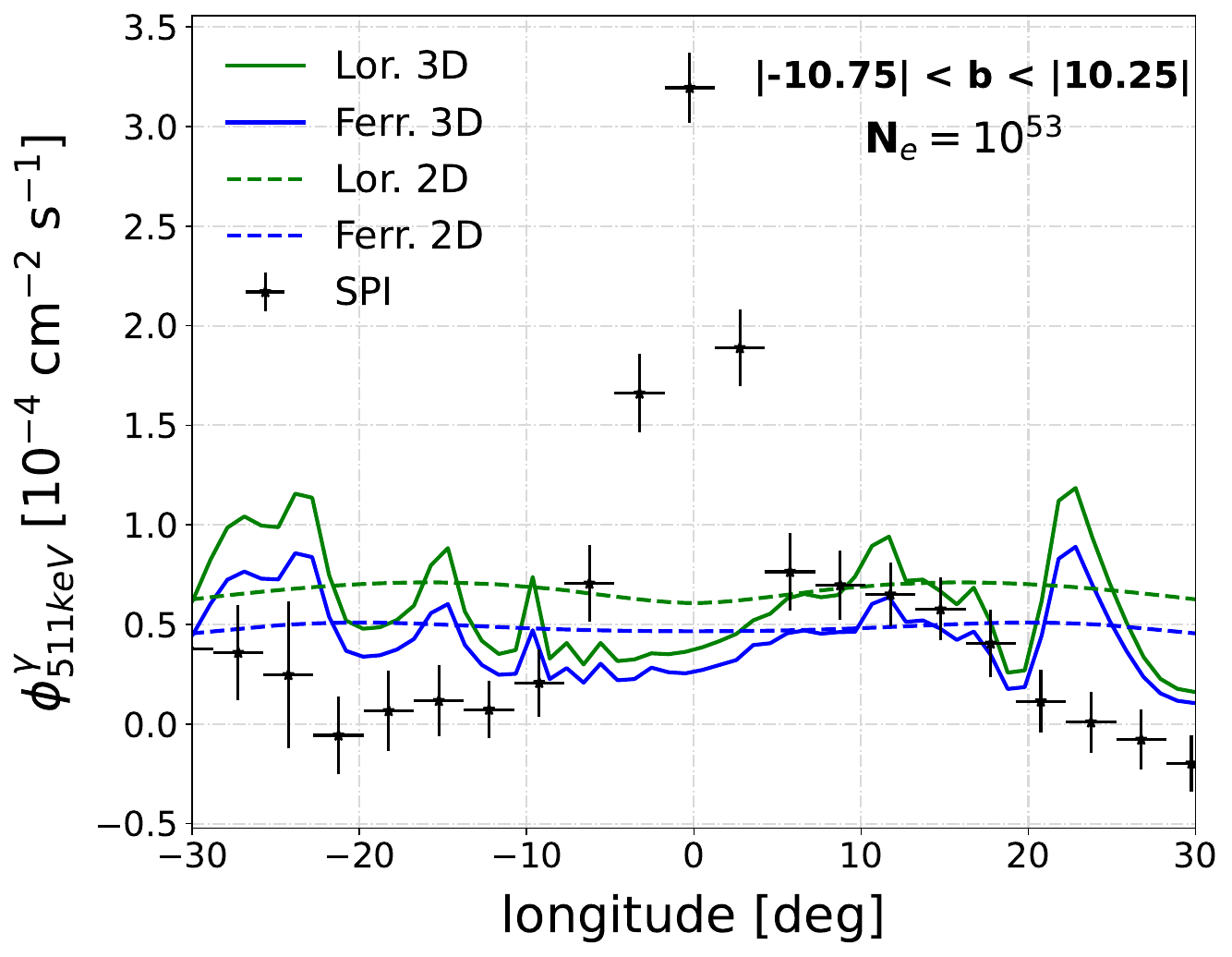}
  \end{minipage} \quad
\caption{{\it Left panel}: Latitude profile for the FIP-induced $511$~keV line signal for our benchmark choice of parameters, $m_{X}=10$~MeV and $N_e =10^{54}$. Observations of SPI are shown in black, with the associated $1\sigma$ experimental error (calculated as the quadrature sum of systematic and statistical errors). Results obtained with the  different SN distributions are shown with different colors, green for Lor and blue for Ferr. The profiles for the 2D distributions, i.e. not considering the spiral arms structure of the Galaxy, are shown as dashed lines. {\it Right panel}: As described in the left panel, but for the longitude profile and $N_e =10^{53}$.}
\label{fig:Main_511}
\end{figure*}
Although obtaining a precise measurement of the galactic diffuse $\gamma$-ray emission at $511$~keV is challenging given the backgrounds involved and the technological limitations, the SPI instrument has provided data collected for more than $20$~years~\cite{Bouchet:2010dj, Siegert:2015knp}. In particular, the latitude and longitude profiles of this emission allow us to constrain the injection of positrons in the Galaxy from different mechanisms. Here, we employ these datasets to derive bounds on the amount of positrons $N_{e}$ injected from electrophilic FIPs, calculated with the approach explained in Sec.~\ref{sec:511calculation}. We also show how these limits are sensitive to important systematic uncertainties and we associate an uncertainty range to the value of $N_{e}$ that can be constrained. We remark that the energy-dependence of this signal is almost independent on the FIP mass, which mildly affects the intensity of the signal for $1~{\rm MeV}\lesssim m_{X}\lesssim20~{\rm MeV}$. Above this mass, the FIP production in SN starts to be Boltzmann suppressed and it is more difficult to achieve a sizable $N_{e}$ in an unconstrained region of the parameter space, depending on the FIP model.
We mention that in our calculations we are also including the continuous emission generated by the same lepton population at $511$~keV, which does not account for more than a $6\%$ of the signal in the considered parameter range.

\begin{figure}[t!]
\includegraphics[width=0.45\textwidth]{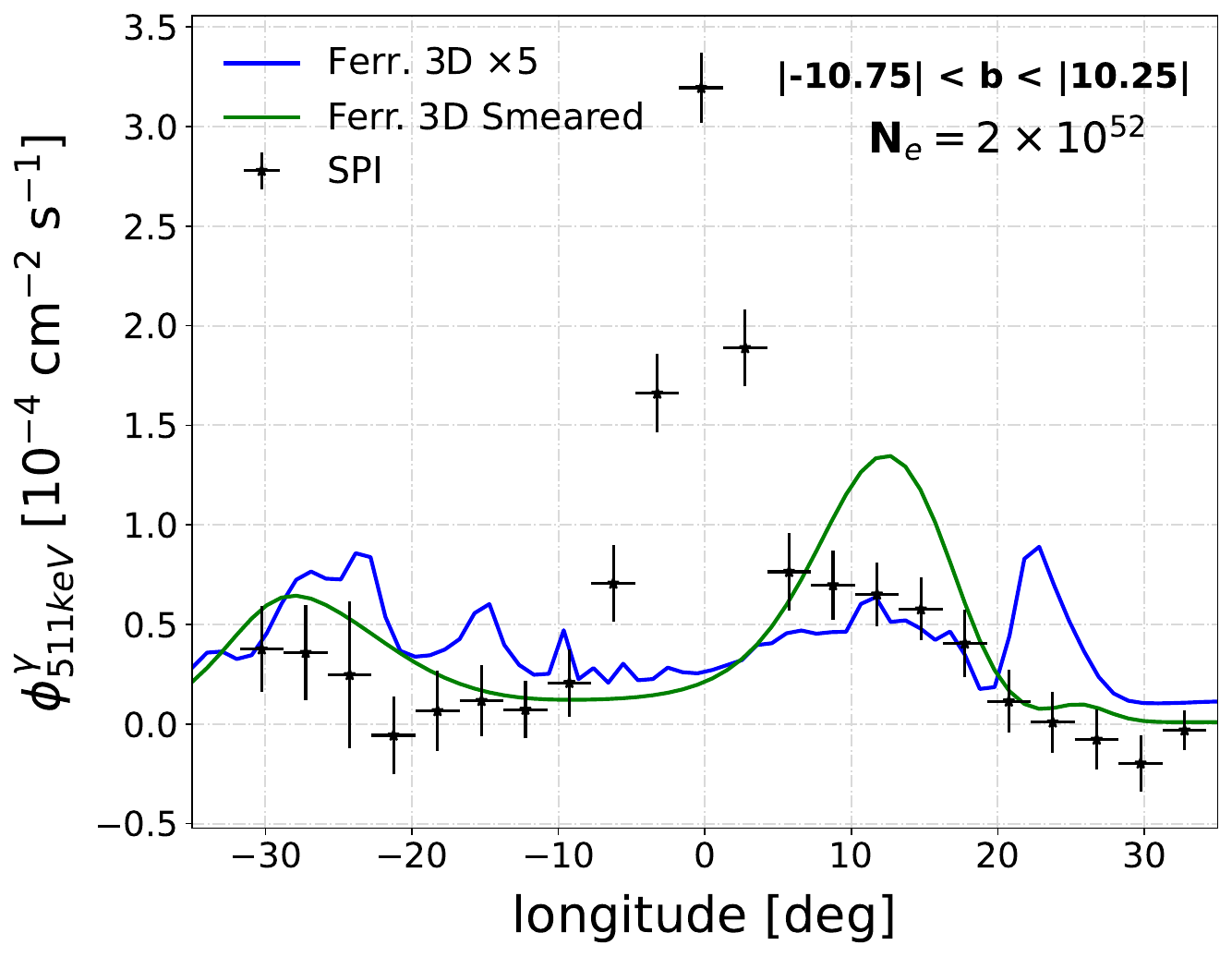}
\caption{Comparison of the predicted $511$~keV longitude profile for the Ferr 3D SN distribution accounting (green line) and not accounting (blue line) for smearing of the signal. The predicted profile without smearing is multiplied by a factor of $5$ to facilitate the comparison. Here, $N_e=2\times 10^{52}$.}
\label{fig:511Smeared}
\end{figure}

In Fig.~\ref{fig:Main_511} we show the expected $511$~keV FIP-induced signal compared to the measured SPI latitude, $N_e =10^{54}$, (left panel) and longitude, $N_{e}=10^{53}$, (right panel) profiles. We show the predicted profiles for different source distributions, modeling the SN distribution as 2D or 3D, when neglecting or including the spiral arms structure of the Galaxy. 
As we show in the left panel, SPI data of the latitude profile roughly follows the behavior of the FIP-induced signal. A radically different situation is shown in the right panel, where the FIP-related signal is peaked at $\sim \pm20^{\circ}$ in longitude, in contrast with measurements peaked towards the galactic center. This is due to the SN distribution, that is not peaked at the galactic center. This different morphology makes it evident that the $511$~keV signal cannot be fitted by the proposed exotic injection of positrons. Hence, this kind of signal is unable to explain the high emission at central longitudes, which is not fully understood in terms of conventional astrophysical mechanisms either~\cite{Siegert:2023wus}.
However, we can set a $95\%$ Confidence Level (CL) upper limit on $N_{e}$, the amount of electrons/positrons produced per SN explosion, as reported in Tab.~\ref{tab:511bounds}. The comparison between limits obtained for different SN distributions allow us to see the impact of this important systematic uncertainty, up to a factor $\sim2$, in our calculations. 
As we see from Tab.~\ref{tab:511bounds} the latitude profile does not give a strict constraint on $N_{e}$, a factor $\sim15$ weaker than the bound from the longitude distribution, reaching $N_{e}\lesssim0.86\times10^{53}$. 
\begin{table}[t!]
\resizebox*{\columnwidth}{0.15\textheight}{
\begin{tabular}{|c|c|c|}
\hline
 \cellcolor{blue!25}\textbf{$N_{e}(\times10^{53})$ $95\%$ CL} & \textbf{Lor} & \textbf{Ferr} \\
\hline
{Latitude profile}  & $12.9$ & $19.7$\\
\hline
{Longitude profile} & $0.86$ &  $1.31$ \\       
\hline
{Long. profile ($|l|<20^{\circ}$)} & $1.52$ &  $2.50$  \\     
\hline
{Smeared Lat. profile}  & $2.60$ & $4.00$\\
\hline
{Smeared Long. profile} & $0.17$ &  $0.20$ \\       
\hline
\end{tabular}
}
\caption{Upper limits at $95\%$ CL on $N_e$ for the Lorimer (Lor) and Ferriere (Ferr) SN spatial distributions for the latitude and longitude profiles of the $511$~keV line~\cite{Siegert:2015knp}. More details are provided in the text.}
\label{tab:511bounds}
\end{table}
The obtained bounds are about one order of magnitude weaker than the upper limit $N_{e}\lesssim1.4\times10^{52}$, obtained in Refs.~\cite{Calore:2021klc,Calore:2021lih}.
This discrepancy can be explained by the more realistic SN distribution used in this work, where we account for the spiral arms structure of the Galaxy. Moreover, up to this point we neglected the smearing effect due to positron propagation before annihilation~\cite{Calore:2021klc,Calore:2021lih}. We also derive limits on $N_e$ taking this effect into account. We employ the same procedure as used in Eq.~(18) of Ref.~\cite{Calore:2021klc} imposing a smearing scale of $1$~kpc. In Fig.~\ref{fig:511Smeared} we show the effect of smearing in the predicted $511$~keV longitude profile for the Ferr SN distribution. As we see, smearing the signal leads to change in the morphology of the profile and allows us to set the most optimistic constraints, as displayed in Tab.~\ref{tab:511bounds}. These limits are a factor of a few better than using our main analysis and are in very good agreement with the results of Refs.~\cite{Calore:2021klc, Calore:2021lih}.

However, in contrast with Refs.~\cite{Calore:2021klc,Calore:2021lih} we find that the smearing effect acts to strengthen the constraint instead of making it weaker. This finding can be understood by examining the right panel of Fig.~\ref{fig:Main_511}. Here, we highlight how the 3D spiral structures suppress the X-ray flux in the most constraining bin at a longitude $l\sim 30^{\circ}$ compared to smooth 2D distributions employed in previous analyses. Thus, roughly speaking, smearing our 3D distributions make the profiles more similar to the 2D ones and the bound is subsequently strengthened. This conclusion is confirmed by the observation that a smearing of the 2D distribution makes the bound weaker, as in Refs.~\cite{Calore:2021klc,Calore:2021lih}.
Therefore, 3D distributions introduce inhomogeneities on very small angular scales, that might weaken the bound, especially at large longitudes, where the most constraining data points are. We do not expect that these structures are observable because of the smearing effect, leading to a stronger constraint. It is important to mention that the approximate smearing of the distribution is a very simple prescription to roughly approximate realistic particle diffusion.

At this point we comment on why constraints obtained with high longitude data might lead to artificially stronger bounds. In Fig.~\ref{fig:Main_511} we observe that some of the data points for $|l|>20^{\circ}$ show negative values of the diffuse flux at $511$~keV.
This is due to the fact that there are important systematic uncertainties in the derivation of these measurements and it is not easy to estimate them for these high-longitude points. Indeed, even the $1\sigma$ upper limit for the most constraining data point at $l\sim 30^{\circ}$ is negative and the bound is set by the $2\sigma$ upper limit. This feature is responsible for making this constraint so stringent.
Issues with the background subtraction at high longitudes can be easily understood if one considers a roughly constant background noise emission: the measurement where the signal is lower would have a lower associated signal-to-noise ratio, leading to more significant uncertainties.
Therefore, in order to avoid the measurements most affected by these systematic uncertainties, we derive conservative limits from the longitude profile, removing large longitude data points in the region $|l|>20^{\circ}$.
These constraints, for the different SN distributions, are shown in the third line of Tab.~\ref{tab:511bounds}, and constitute the most conservative ones, being around a factor $\sim2$ weaker than the non-smeared constraints previously discussed.
This analysis allows us to associate uncertainty bands on the  $511$~keV bound and subsequently $N_{e}$. More precisely, the most stringent bound is obtained by high longitude data points including smearing effects and the 3D Lor SN profile
\begin{equation}
    N_{e}\lesssim 0.17\times10^{53}\,.
\end{equation}
The weakest constraint is obtained by conservatively neglecting smearing effects, considering only low-longitude data points and the 3D Ferr SN distributions
\begin{equation}
    N_{e}\lesssim 2.50\times10^{53}\,.
\end{equation}
Therefore we can conservatively associate more than one order of magnitude of uncertainty to this constraint.  
On top of this, we remark on the fact that using this approximate evaluation of the 511 keV line leads to profiles that are insensitive to the energy dependence of the positron spectrum. This is not completely physical, since particles of different energy will propagate at different rates, implying that their distribution will correspondingly vary. This is another caveat that adds to the uncertainty mentioned above.

\subsection{Local electron/positron measurements}
\label{sec:voyAMS}

\begin{figure}[t!]
\includegraphics[width=\columnwidth]{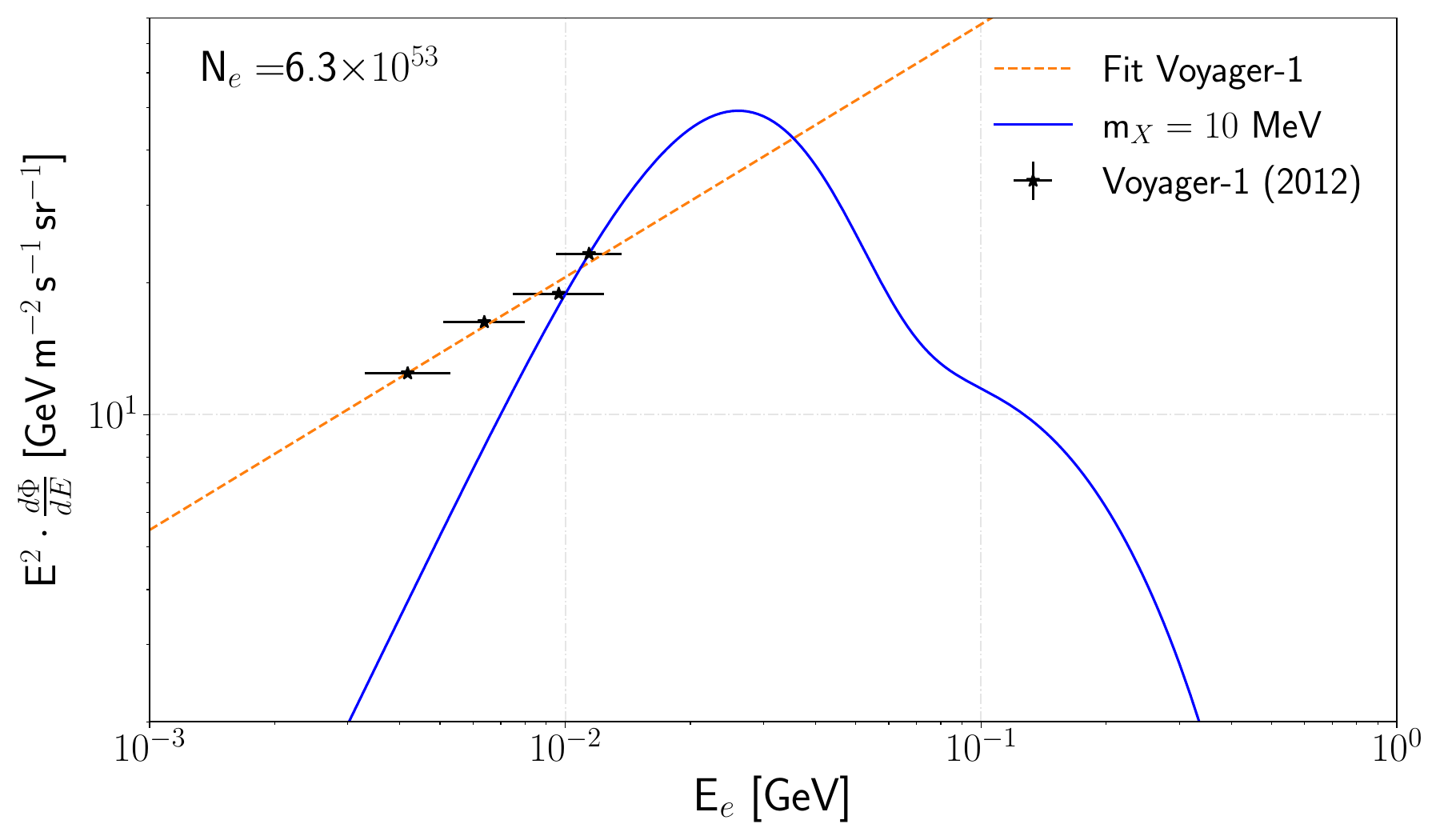}
\caption{Comparison of the $e^\pm$ flux measured by Voyager-1 with the FIP-induced signal for $N_e=6.5 \times 10^{53}$ and $m_{X}=10$~MeV. We include, as an orange dashed line, the result of a power-law fit to this data for completeness (see the text for more details).}
\label{fig:VoyFit}
\end{figure}

Electrons generated in electrophilic FIP decays would contribute to the galactic diffuse electron population at MeV energies. This allows us to use Voyager-1 measurements of the local electron flux to constrain their injection. 
For a simple order of magnitude estimate, we can evaluate the flux of injected electrons in the Galaxy as
\begin{equation}
 \frac{d\Phi_e}{dE_e}\simeq\frac{\Gamma_{cc}}{4\pi d_{SN}^2}  \frac{N_{e}}{\Delta E_{e}} \,,
\end{equation}
where $\Delta E_{e}\simeq 50$~MeV is the energy range over which electrons are injected. Comparing this number with the electron flux measured by Voyager-1 in Fig.~\ref{fig:VoyFit} (black points), we realize that the FIP-induced flux leaves an imprint on this observable if $N_{e}\simeq10^{54}$.
This expectation is confirmed by the full numerical evaluation of the electron flux (blue line), shown in Fig.~\ref{fig:VoyFit}. This is compared to the Voyager-1 measurements in the $4$-$10$~MeV range. Since the Voyager-1 data correspond to the sum of $e^+$ and $e^-$ i.e. $e^\pm$, our limit is obtained considering the sum of both species.
Here, we have added a line obtained from a power-law fit of the Voyager-1 data that allows us to illustrate how our limit on $N_e$ could improve (up to a factor of $3$) if data at slightly higher energies would be available. 
We notice that the uncertainties related to determining the propagation parameters play a minor role in this energy range. However, we observe that the main uncertainty could be the halo height employed, for which these signals scale as the square root (the larger the halo height, the larger the signal). Hence the limits derived from the Voyager-1 data have an associated systematic uncertainty of less than $50\%$, corresponding to extreme values of the halo height in the range $3$-$20$~kpc. 

\begin{figure*}[t!]
  \begin{minipage}{.45\textwidth}
    \includegraphics[width=\linewidth]{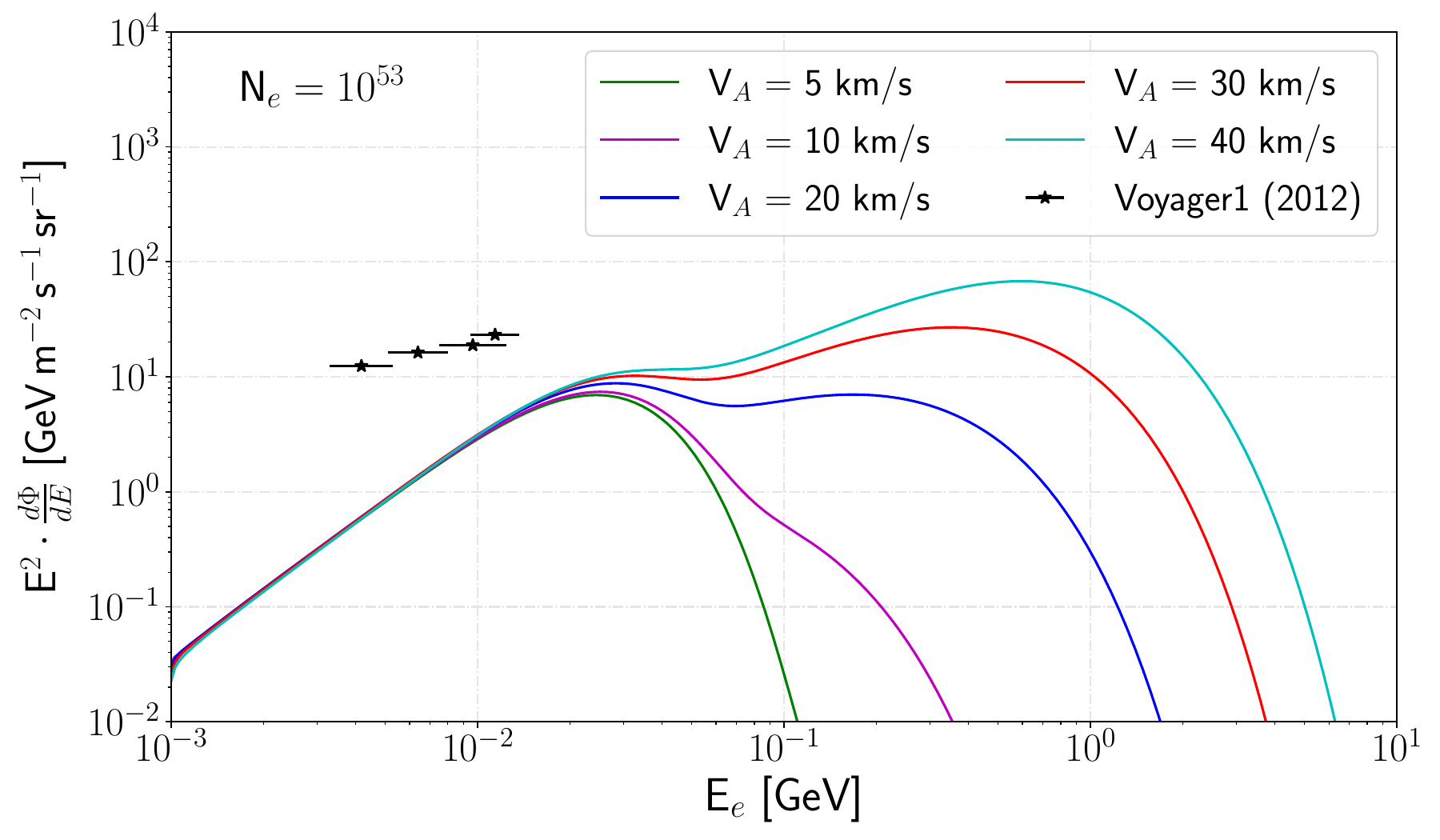}
  \end{minipage} \quad
  \begin{minipage}{.45\textwidth}
    \includegraphics[width=\linewidth]{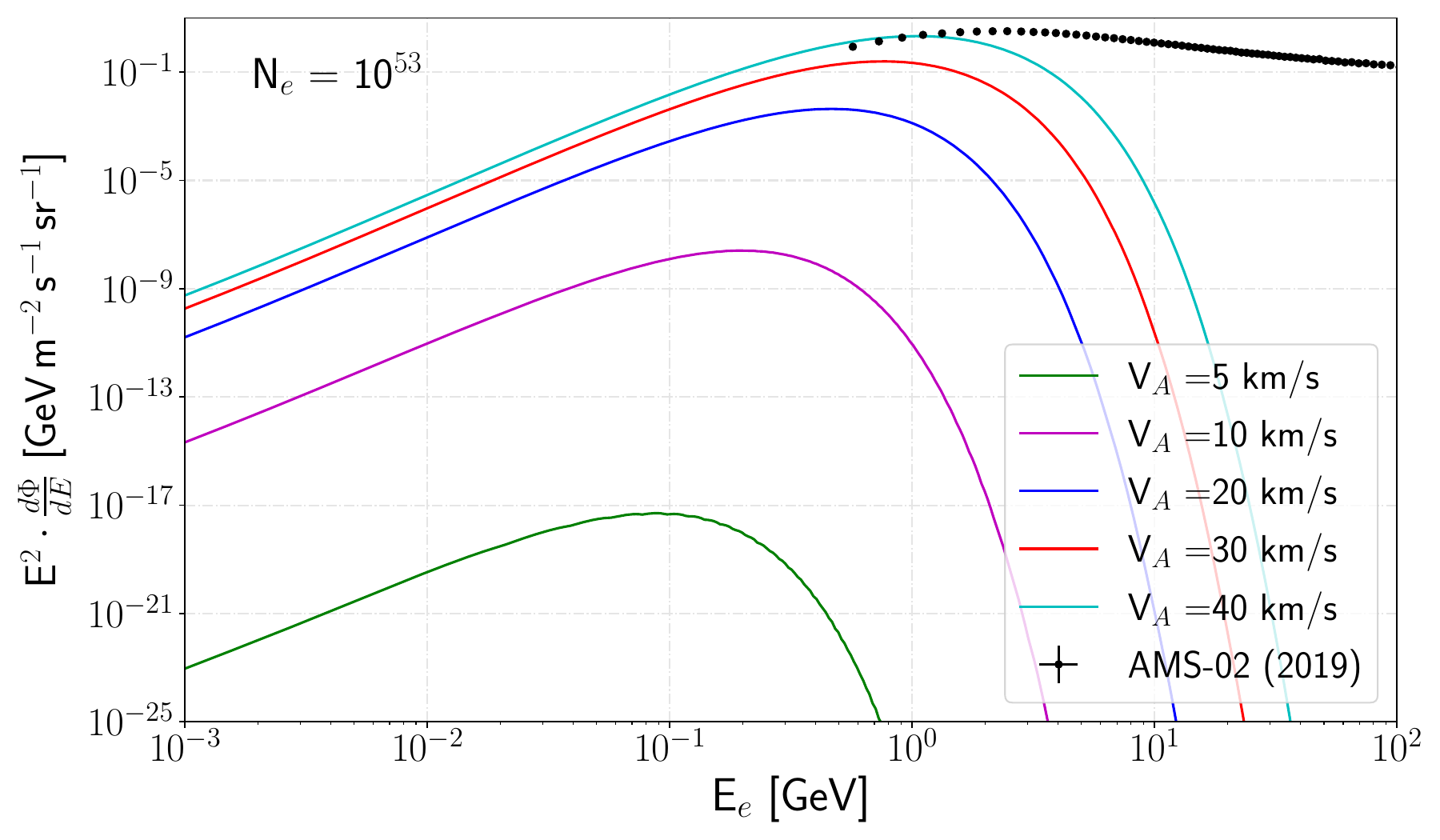}
  \end{minipage} \quad
\caption{{\it Left panel:} Effect of reacceleration on the predicted FIP-induced electron signal for $N_e=10^{53}$ and $m_{X}=10$~MeV.
Comparison of the Voyager-1 $e^\pm$ flux data with the FIP-induced electron spectrum for different values of the Alfvèn velocity (V$_A$). {\it Right  panel:} Similar to the left panel, but for positron Solar-modulated spectra compared to the AMS-02 positron data.}
\label{fig:Reacc}
\end{figure*}

The left panel of Fig.~\ref{fig:Reacc} aids in illustrating the weak dependence of the Voyager-1 bound to systematic uncertainties. Precisely, this comparison shows the effect of reacceleration.
At energies above $\sim100$~MeV the effect of reacceleration in the spectrum of the FIP-induced electron population becomes dramatic. It basically implies a gain of energy for low-energy CR particles due to the interaction with magnetic plasma waves. The effect of reacceleration is proportional to the square of the speed of the plasma waves, i.e. the Alfvèn velocity and inversely proportional to the spatial diffusion coefficient.
In conclusion, different Alfvèn velocities do not affect the limits from Voyager-1, since the electron spectrum remains roughly unchanged below $\sim20$~MeV, while these fluxes are enhanced by orders of magnitude above $100$~MeV depending on the exact value of this parameter.

In principle, for high reacceleration, we could derive constraints on $N_e$ from positron AMS-02 data at GeV energies (see the right panel of Fig.~\ref{fig:Reacc}). The solar-modulated spectrum of the FIP signals for different values of the Alfvèn velocity, from $5$~km/s to $40$~km/s is shown. Because of the lower background production of positrons, the positron channel would provide stronger constraints than the electron flux at those energies.
Remarkably, we observe that for large enough values of the Alfvèn velocity ($V_{A}\sim 30$-$40\km/\s$) we would be able to obtain strong constraints from AMS-02. As opposed to the Voyager-1 data, that are obtained from outside the heliosphere, AMS-02 is located near Earth, meaning that the particles that it measures are affected by the heliosphere and the solar magnetic fields. Unfortunately, the interaction with the heliosphere introduces a strong suppression in the flux of these particles for small reacceleration.
Analyses of secondary CRs usually favor values of the Alfvèn velocity between a few km/s and $\sim30$~km/s~\cite{Luque:2021nxb}, so values above this would be in conflict with CR observations. As a matter of fact, our benchmark value is $13$~km/s, which yields a FIP-induced signal many orders of magnitude below AMS-02 data. The main reason that Voyager-1 data can provide stronger constraints than AMS-02, is the suppression of the flux of CR particles from their interaction with the heliosphere. Regarding positron fluxes, we cannot set constraints on FIPs because of the strong variability of the signal with the Alfvèn velocity uncertainties. In the case of maximal reacceleration the constraint would be comparable with the Voyager-1 bound.

In conclusion, we observe that Voyager-1 provides very valuable observations to constrain electrophilic FIPs at MeV masses. In particular, the limits derived from Voyager-1 data ($e^+$ + $e^-$) are of the order of $N_e\simeq 6$-$9\times 10^{53}$, which are compatible with those found from the analysis of the SPI profiles. Additionally, we remark that these bounds are much less affected by the different sources of systematic uncertainties present in the modeling of the $511$~keV signal.

\subsection{X-to-$\gamma$-ray data}
\label{sec:XGamma}

If electrophilic FIPs produce electrons and positrons with energies above tens of MeV, these can emit high-energy photons via IC and bremsstrahlung. This flux is probed by diffuse $\gamma$-ray observations from keV to MeV energies in various galactic regions.
The quality of data in this energy range is not as high as it is at the GeV scale, mainly due to the high instrumental backgrounds. We have collected a variety of valuable measurements and show that by using different observational data, referring to different energy ranges and angular positions in the sky, we can set bounds on FIPs competitive with ones arising from the $511$~keV line and direct observations of electrons.
In particular, we use the observations from the following experiments:
\begin{itemize}
    \item EGRET~\cite{Strong:2004de} measurements at energies higher than $20$~MeV and above the galactic plane,
    \item COMPTEL~\cite{1998PhDT.........3K} data in a region around the center of the Galaxy in the range $2$-$20$~MeV,
    \item SPI~\cite{Berteaud_2022, Siegert:2022jii, Bouchet:2008rp} data between $20$~keV and $2$~MeV for two sectors of the sky around the galactic center,
    \item MOS~\cite{Foster:2021ngm} (in the XMM-Newton mission) observations from $2.5$ to $8$~keV in several angular rings around the center of the Galaxy. 
\end{itemize}
With the combination of these datasets we cover a wide energy range, from keV to tens of MeV, for different regions of the sky, making a comprehensive evaluation of the FIP constraints possible. In all the cases we consider the full uncertainties (statistical and systematic errors) reported by the experiments and obtain limits at $95\%$ CL from a $\chi^2$ analysis. The obtained limits are shown in Tab.~\ref{tab:XGbounds}. In the following, we discuss the results from each dataset. 

\begin{table}[t!]
\resizebox*{0.8\columnwidth}{0.15\textheight}{
\begin{tabular}{|c|c|}
\hline
 &\cellcolor{blue!25}\textbf{$N_{e}(\times10^{55})$ $95\%$ CL}  \\ 
\hline
{EGRET}  &  $2.2$ \\  
\hline
{COMPTEL}  &  $0.18$ \\ 
\hline
{SPI ($|l| = |b| < 47.5^{\circ}$)} & $2.5$ \\    
\hline
{SPI ($|b|<15^{\circ}$, $|l|<30^{\circ}$) }  & $2.1$ \\     
\hline
{MOS/XMM (ring 1)}  & $0.35$  \\
\hline
{MOS/XMM (ring 3)} & $0.047$  \\       
\hline
{MOS/XMM (ring 4)} & $0.06$  \\       
\hline
{MOS/XMM (ring 30)} & $10.1$  \\       
\hline
{MOS/XMM (Combined)} & $0.027$  \\       
\hline
\end{tabular}
}
\caption{Limits on $N_e$ at the $95\%$ CL derived from the fit of the predicted secondary X-to-$\gamma$-ray emissions to the datasets discussed in the text.}
\label{tab:XGbounds}
\end{table}

\begin{figure*}[t!]
  \begin{minipage}{.45\textwidth}
    \includegraphics[width=\linewidth]{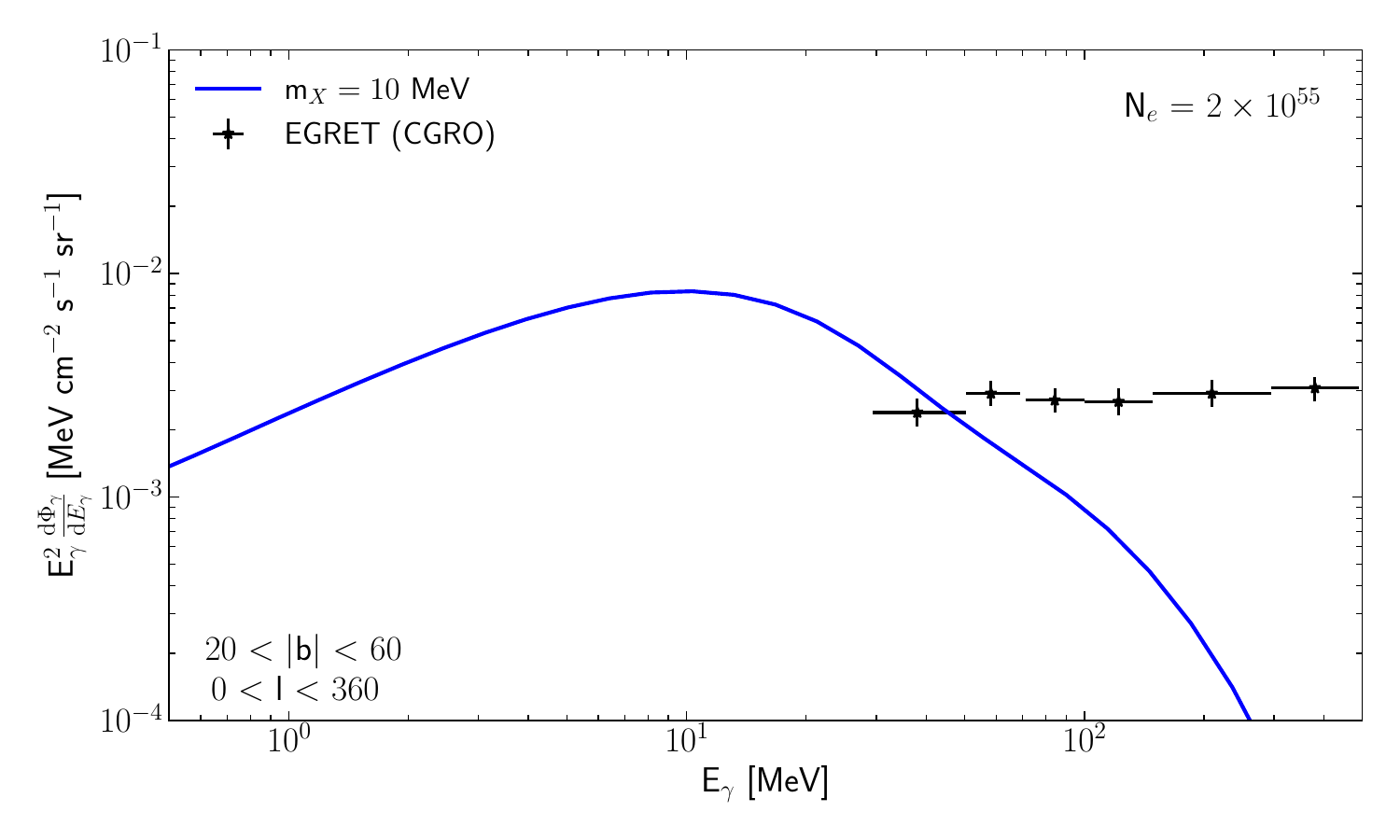}
  \end{minipage} \quad
  \begin{minipage}{.45\textwidth}
    \includegraphics[width=\linewidth]{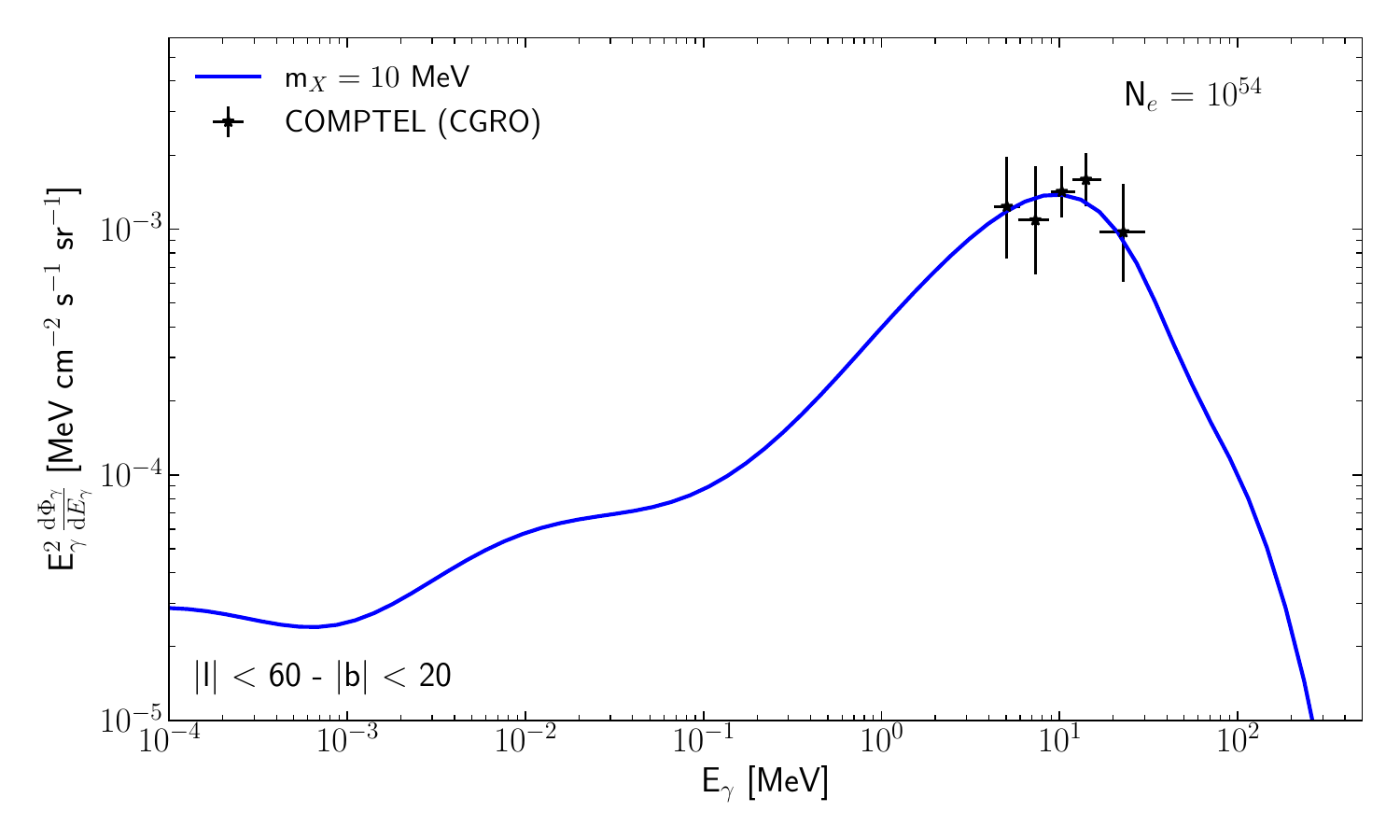} 
  \end{minipage} \quad

  \begin{minipage}{.45\textwidth}
    \includegraphics[width=\linewidth]{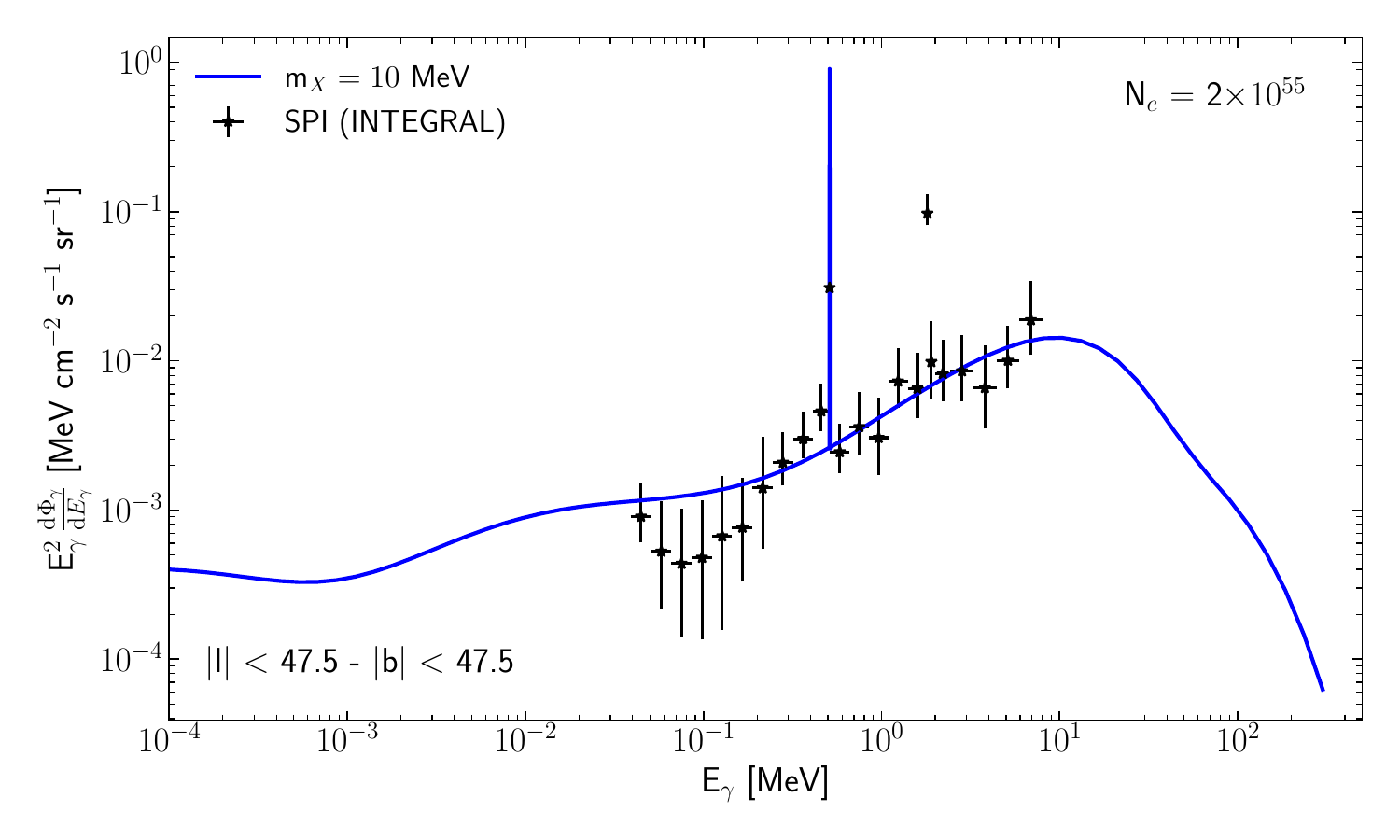} 
  \end{minipage} \quad
  \begin{minipage}{.45\textwidth}
    \includegraphics[width=\linewidth]{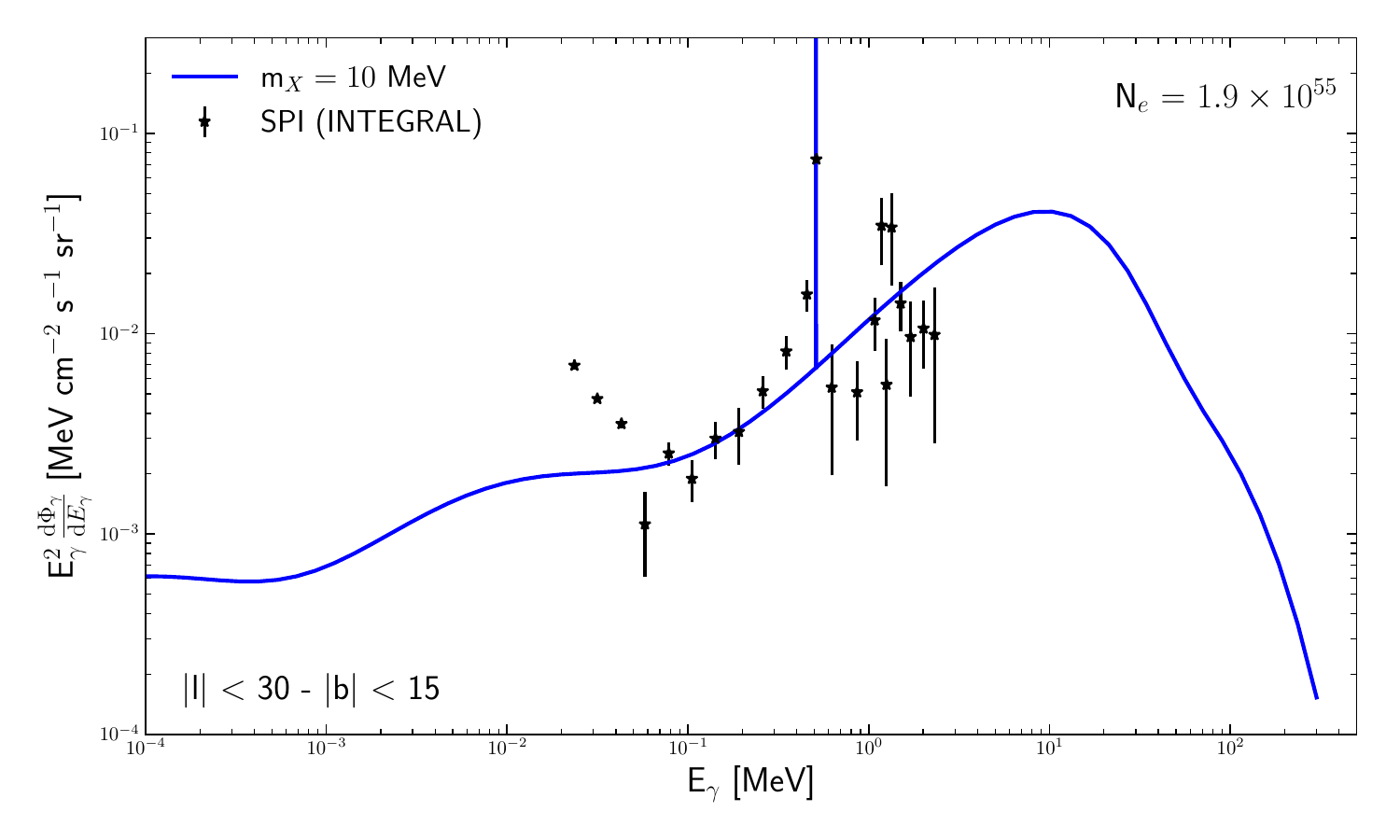} 
  \end{minipage} \quad
   \begin{minipage}{.45\textwidth}
    \includegraphics[width=\linewidth]{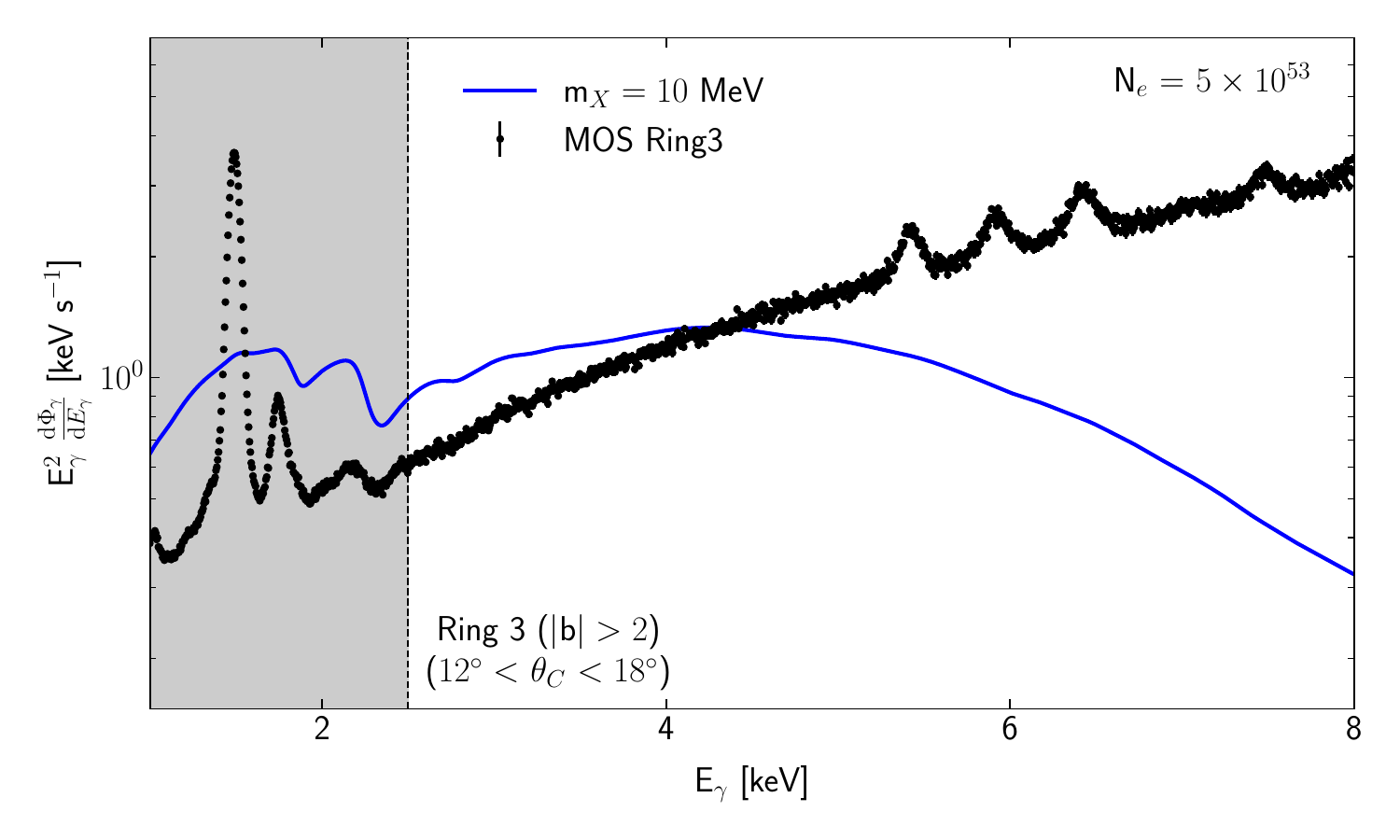}
  \end{minipage} \quad
  \begin{minipage}{.45\textwidth}
    \includegraphics[width=\linewidth]{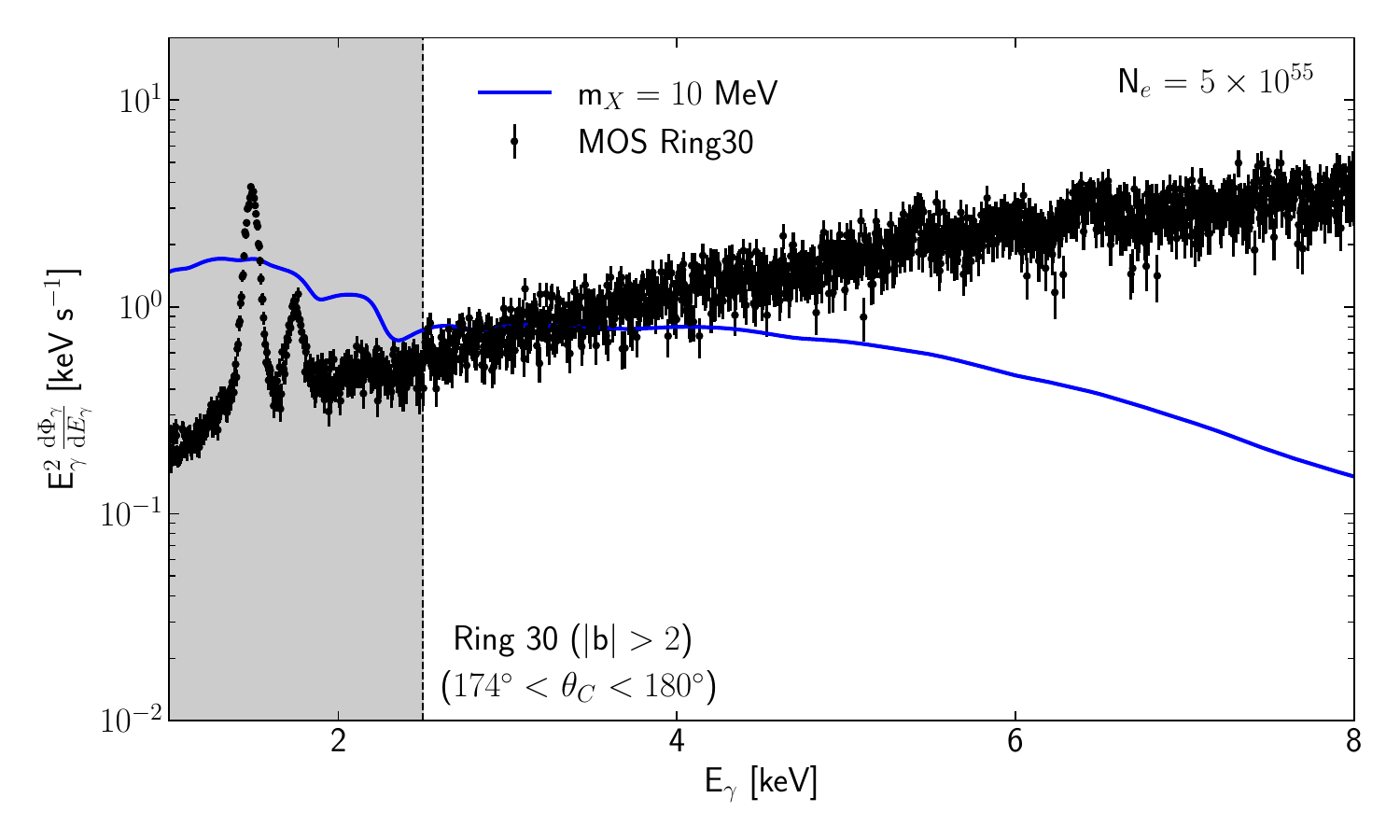} 
  \end{minipage} \quad
\caption{Comparison of different measurements with the predicted secondary signal for $m_{X}=10$~MeV and evaluated in the region of the sky where the the data is extracted in each case. We show the total $1\sigma$ error bars reported by the experiments.
{\it Top-left panel:} Comparison with EGRET (CGRO) data, extracted from Ref.~\cite{Strong:2004de}. {\it Top-right panel:} COMPTEL (CGRO)~\cite{1998PhDT.........3K}.
{\it Middle-left panel:} SPI (Integral)~\cite{Berteaud_2022}.  {\it Middle-right panel:} SPI (Integral)~\cite{Bouchet:2008rp}.  
{\it Bottom-left panel:} MOS data at the Ring 3~\cite{Foster:2021ngm}. We shade the region not used in the fits to this data since we consider only the range $2.5$-$8$~keV. {\it Bottom-right panel:} MOS data at Ring 30~\cite{Foster:2021ngm}. }
\label{fig:MeVs}
\end{figure*}

\textbf{Constraints from EGRET data}: In the upper left panel of Fig.~\ref{fig:MeVs}, we show the comparison of the FIP-induced signals expected for $N_e=2\times 10^{55}$ and $m_{X}=10$~MeV with the data extracted from Ref.~\cite{Strong:2004de}, for the region of the Galaxy at $20^{\circ}<|b|<60^{\circ}$ and $0^{\circ}<l< 360^{\circ}$. In this case, the peak of the bremsstrahlung emission, at $E_{\gamma}\sim10$~MeV, lies at energies below the available EGRET data and, therefore, the signal's flux falls down rapidly in the EGRET measurements energy range, leading to weak constraints in comparison to those derived from the $511$~keV line or Voyager-1 data. However, the lowest-energy data points allow us to still derive relevant constraints from this dataset.

\textbf{Constraints from COMPTEL data}: In the upper right panel of Fig.~\ref{fig:MeVs}, we show how the expected FIP signals compare to COMPTEL data~\cite{1998PhDT.........3K} for the region $|b|<20^{\circ}$ and $|l|<60^{\circ}$. Here, we only use the data points above the region of the emission lines at $2.23$~MeV, produced from the formation of deuterium in the interstellar medium, and $\simeq 1.8$~MeV where lines are caused by the decay of $^{26}$Al into $^{26}$Mg and its subsequent de-excitation. We are not including these sources of background, whose modeling is far beyond the scope of this work. Since the bremsstrahlung peak emission coincides with the position of these data points, strong constraints can be derived using this dataset, competitive to those from the $511$~keV SPI latitude profiles and those from Voyager-1 data. In particular, we obtain an upper limit in the number of electrons/positrons injected of $N_e\sim1.8\times10^{54}$ at $95\%$ CL (see Tab.~\ref{tab:XGbounds}).

\textbf{Constraints from SPI data}: In the middle left panel of Fig.~\ref{fig:MeVs} we show a comparison of the FIP-induced signal with the data of the diffuse $\gamma$-ray emission recently derived in Ref.~\cite{Berteaud_2022} 
for a region with latitude and longitude smaller than $47.5^{\circ}$ around the galactic center. This dataset lies slightly below the bremsstrahlung peak emission and leads to very weak constraints on the injected number of electrons and positrons from electrophilic FIPs when compared to the other limits derived in this work. 

In the middle right panel of Fig.~\ref{fig:MeVs}, we show the SPI data obtained by Ref.~\cite{Bouchet:2008rp} at $|b|<15^{\circ}$ and $|l|<30^{\circ}$. Note that this is a region very concentrated around the galactic center, leading to slightly stronger constraints than in the previous case. This dataset covers the energy region from $\sim20$~keV to $\sim2$~MeV, where the FIP emission transitions from being dominated by the bremsstrahlung emission to being dominated by the IC emission (below $\sim1$~MeV).    
In this figure, for illustrative purposes we show the associated line emission at $511$~keV from positronium annihilation. Although we remark that the constraint on $N_e$ from this dataset is derived using the continuous emission. As we see, the line emission exceeds by far the measurements at $511$~keV for electron injection values which are not ruled out by SPI continuous emission. These limits are shown in Tab.~\ref{tab:XGbounds}.

\begin{figure*}[t!]
\begin{minipage}{\textwidth}
\includegraphics[width=0.95\textwidth]{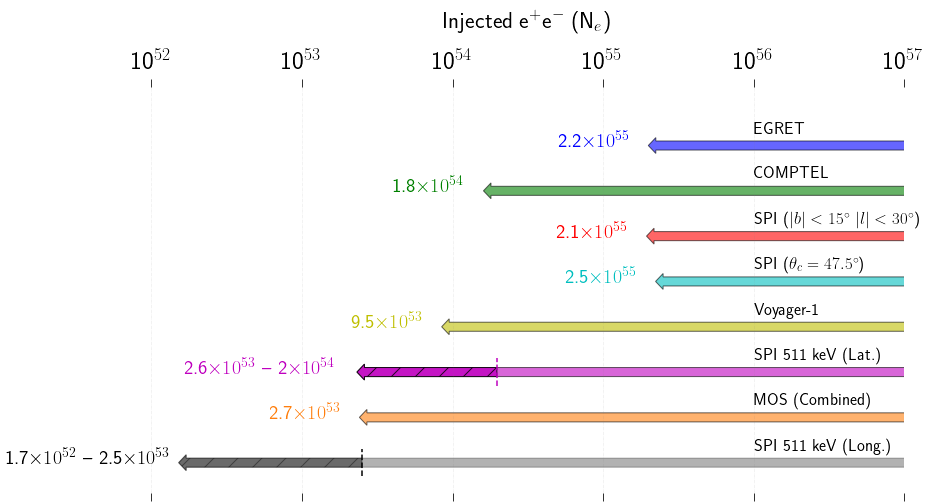}
\end{minipage} 
\caption{Representation of the limits on $N_e$ at the $95\%$ CL derived from the fit of the predicted secondary X-to-$\gamma$-ray emissions to the datasets discussed in the text. The discussed uncertainties in the evaluation of the limits from the profiles of the $511$~keV emission are shown as hatched bars. These limits are valid for FIP masses below $\sim20$~MeV.}
\label{fig:Limits_Arrow}
\end{figure*}

\textbf{Constraints from XMM data}: We use the data from the MOS detector in the energy range from $1$~keV to above $10$~keV, which was provided by Ref.~\cite{Foster:2021ngm} divided in galactocentric rings that are $6^{\circ}$ wide. In this case, the most central rings lead to a higher FIP-induced signal since the density of SN tends to increase towards the galactic center. We observe that this is the trend that our constraints follow, being stronger for the inner rings and weaker in the external rings. However, the limits derived from the two innermost rings (Rings 1 and 2) are weaker than the limit derived from Ring 3 (the strongest limit), as it was observed in Ref.~\cite{Cirelli:2023tnx}, which is likely due to the very high astrophysical backgrounds and lower FIP-induced signals that are present in the very central regions.
Here, we show the comparison of the calculated  signals with MOS data at the most constraining ring (Ring 3) and the most external ring (Ring 30) in the lowest row of Fig.~\ref{fig:MeVs}. We observe that, while an injection of electrons of $N_e=5 \times 10^{53}$ is already above MOS data in the $2$-to-$4$~keV range for Ring 3 (left panel), the most external ring only allow us to constrain $N_e$ to be smaller than $\sim1\times 10^{56}$.
In particular, the $95\%$ CL limits that we show in Tab.~\ref{tab:XGbounds} are derived with the data from $2.5$ to $8$~keV, as performed by Refs.~\cite{Foster:2021ngm,Cirelli:2023tnx}, due to the background noise in the detector at lower and higher energies respectively. We follow a conservative approach in our analysis by omitting energy regimes in the MOS detector data that are dominated by instrumental noise. In addition to this, the combination of all the rings allows us to set a constraint that is a factor 2 stronger. Remarkably, these are the strongest limits derived from the continuous X-to-$\gamma$-ray emission. They are notably even stronger than those from the Voyager-1 electrons, and competitive with those of the longitude profile of the $511$~keV line, which were the strongest imposed so far. This, added to the fact that MOS measurements and its uncertainties are much more robust than those evaluated for the $511$~keV line, makes the MOS instrument exceptionally valuable for constraining FIPs. 
We have tested how these constraints are affected by different values of $E_0$, finding that they only change by a factor of $2$ with respect to the case with $E_0=30$~MeV and with $E_0=60$~MeV.
In addition, we remark that higher reacceleration values, improve these limits by a factor of a few, making the MOS data even more valuable for constraining electrophilic FIP signals.

\textbf{Summary of the constraints}: In Fig.~\ref{fig:Limits_Arrow}, we summarize the limits obtained on the number of electrons/positrons injected by FIPs with $m_{X}=10$~MeV. We expect these constraints to be valid for the entire $1$-$20$~MeV range of FIP masses. We also show hatched regions for the limits extracted from the emission profile of the $511$~keV line, representing the uncertainties that we have estimated and propagated. As we mentioned earlier, the evaluation of the 511 keV line does not include the full propagation, energy losses and other relevant effects for the positrons, this leads to a correspondingly crude estimate for the FIP constraints that result from its dataset.
Notably, the data from the MOS detector leads to constraints compatible with the lower limit obtained from the longitude profile of the $511$~keV line. We remark that, as we commented above, SPI data seems to be largely affected by systematic uncertainties not well accounted for, while the measurements by the MOS detector seem to be very robust, leading to more reliable constraints. Similar reasoning applies for theoretical uncertainties on the calculation of these fluxes. Additionally, we also notice that Voyager-1 data provides constraints that are compatible with those from the latitude profile of the $511$~keV line. Among the $\gamma$-ray datasets used here, the measurements from COMPTEL provide the strongest constraints. Other uncertainties related to the modelling of the propagation of sub-GeV electrons can affect these limits, depending on the dataset and the observable employed. However, a full evaluation of these uncertainties is beyond the scope of this work.

\begin{table*}[t!]
    \centering
    \begin{tabular}{c|c|c|c|c}
      &ALPs&\hspace{0.2cm}DPs\hspace{0.2cm}   & \hspace{0.2cm}$\nu_{s}-\mu s$\hspace{0.2cm} &\hspace{0.2cm}$\nu_{s}-\tau s$\hspace{0.2cm}  \\
      \hline
       $\alpha$  &0 &2&1.64&1.52\\
       $N_{0}$ &$5.2\times10^{56}$ &$1.5\times10^{57}$&$1.3\times10^{59}$&$1.3\times10^{59}$\\
       $g_{0}$  & $g_{ae}=2.7\times10^{-13}$&$\epsilon=7.9\times10^{-11}$&$|U_{\mu s}|=1.8\times10^{-2}$&$|U_{\tau s}|=0.11$\\
     $\tau(g_{0})$ &0.14&3.06 &0.004&0.37\\
    \end{tabular}
    \caption{Parameters of Eq.~\eqref{eq:fitting} to fit $N_{e}$ as function of the FIP coupling for ALPs, DPs and sterile neutrinos mixed with muon or tau flavors. }
    \label{tab:my_label}
\end{table*}

\section{Application to FIP constraints}
\label{sec:FIPconstr}

The upper limits on the electron/positron injection discussed in Sec.~\ref{sec:constraints} can be applied to several FIP models. In the literature, some interest was devoted towards constraining sterile neutrino properties~\cite{Dar:1986wb,Calore:2021lih}, ALPs~\cite{Calore:2021klc} and DPs~\cite{Calore:2021lih}. The same models can be studied in the light of the complementary constraints introduced in this work. If these particles exist in nature, we expect them to be produced in SN and possibly have a coupling with electrons/positrons. Even though accurately revisiting all these constraints is beyond the scope of this work, we will briefly sketch how the existing constraints improve and become more robust.

In a very general fashion, we can write down the number of positrons produced by FIPs in a single SN explosion as function of the FIP-electron coupling $g$
\begin{equation}
    \begin{split}
        N_{e}=N_{0}\left(\frac{g}{g_{0}}\right)^{\alpha}e^{-\left(\frac{g}{g_{0}}\right)^{2}\tau(g_{0})}\,,
    \end{split}
    \label{eq:fitting}
\end{equation}
where $N_{0}$, $g_{0}$ and $\tau(g_{0})$ are parameters fitting various physical situations. The parameter $\alpha=\{0,2\}$ represents the possibility of having production of FIPs and their decay controlled by the same coupling $g$ or not. Namely, as discussed in Ref.~\cite{Calore:2021klc} a copious amount of positrons is produced by ALPs emitted by a SN via their coupling to protons and decaying via their electron-interaction $g=g_{ae}$. In this case, since production and decay are caused by different couplings, $\alpha=0$. Otherwise, in the most common case, both decay and production are proportional to the same coupling and $\alpha=2$, like for DPs, mixed with photons with a coupling $g=\epsilon$ and sterile neutrinos mixed with active ones through $g=|U_{\alpha s}|$ for $\alpha=\mu,\tau$~\cite{Calore:2021lih}. In the case of sterile neutrinos $\alpha$ can assume a value different from $2$ in order to schematically take into account the contribution of type Ib/c SN to the positron production.
The behavior of the fitting function in Eq.~\eqref{eq:fitting} can be understood by looking at the extreme regimes of free-streaming, $(g/g_{0})^{2}\tau(g_{0})\ll1$, and trapping $(g/g_{0})^{2}\tau(g_{0})\gg 1$. In the first case, for $\alpha=2$, the positron number increases proportional to $g^2$, since the FIP production increases. In the opposite limit, the exponential suppression is severe because of FIPs decaying inside the SN envelope and no positrons being able to escape the star.  In these terms, the requirement that FIPs decay close to the SN translates into a lower limit on the FIP opacity
\begin{equation}
   \left(\frac{g}{g_{0}}\right)^{2}\tau(g_{0})\gtrsim \frac{R_{*}}{R_{\rm dec}}\simeq3\times10^{-11}\,,
   \label{eq:lowerlimit}
\end{equation}
where $R_{*}\simeq 10^{7}$~km is the SN (Ib/c) envelope radius and $R_{\rm dec}\lesssim10$~kpc is the average distance between SN. 

We use Eq.~\eqref{eq:fitting} to fit the positron production from ALPs (with $g_{ae}\gtrsim10^{-14}$), DPs and sterile neutrinos, all the particles having a mass of 10~MeV. We summarize the fitting parameters in Tab.~\ref{tab:my_label}, based on the numerical calculations presented in Refs.~\cite{Calore:2021klc,Calore:2021lih,Carenza:2023old}.
It is easy to verify that, with the reported parameters, the condition in Eq.~\eqref{eq:lowerlimit} is always satisfied for the parameter range considered. Therefore, for most of the considered FIP models, except for ALPs with weak coupling to electrons, our assumption that the positron injection happens close to the SN is confirmed. 
Now we are in a position to estimate the exclusion regions for the various FIP parameters, given the phenomenology discussed in this work. From Fig.~\ref{fig:Limits_Arrow}  we notice that the SPI constraint on the 511~keV line longitudinal profile is the strongest constraint on the electron-positron injection from SN, reaching 
\begin{equation}
    N_{e}\lesssim1.7\times10^{52}\,,
\end{equation}
in agreement with Refs.~\cite{Calore:2021lih,Calore:2021klc}. In this work we showed that this constraint, even though it is the strongest one, is affected by several theoretical uncertainties related to the SN distribution, choice of the most constraining dataset and modeling of the positron propagation in the Galaxy (see Sec.~\ref{sec:511}). These uncertainties may relax the bound by up to one order of magnitude. We propose a more robust bound set by MOS observations of the IC signal at a few keV caused by the leptons injected by FIP decays.  With these observations we are able of setting a very robust constraint of
\begin{equation}
    N_{e}\lesssim2.7\times10^{53}\,.
\end{equation}
A summary of the various constraints translated to FIP couplings are shown in Fig.~\ref{fig:FIPbound}. Note that these are only rough estimates of the constraints because we used the approximate fitting function in Eq.~\eqref{eq:fitting} and not the full numerical approach, approximating the non-trivial effect of type Ib/c SN~\cite{Calore:2021lih}. For a first estimate, we expect this approach to be reasonably accurate and especially useful to give an idea of how the different constraints reflect on FIP properties. As expected, the MOS bound is comparable with the 511~keV one, giving the most stringent and robust constraint on $N_{e}$. In addition, the Voyager-1 and COMPTEL constraints help to set a slightly weaker constraint. This combination of data allows us to robustly exclude a large portion of the various FIPs parameter space {beyond the portion already excluded by other astrophysical arguments, such as radiative decay of the various FIPs escaping the SN, or depositing energy inside the envelope~\cite{Sung:2019xie,Calore:2021lih,Carenza:2023old} (red hatched regions). This shows that the different observables employed in this work lead to a family of independent, competitive and very strong limits on FIP production. In the case of ALPs, we are able to set the only constraints in this region of the parameter space, which is why there is no hatched region from other constraints shown in Fig.~\ref{fig:FIPbound}. Given the relevance of these bounds for FIP searches, we leave a more accurate evaluation of the various constraints for FIPs of different masses for a future work~\cite{Balajinew}.

\begin{figure*}[t!]
\begin{minipage}{\textwidth}
\includegraphics[width=0.99\textwidth]{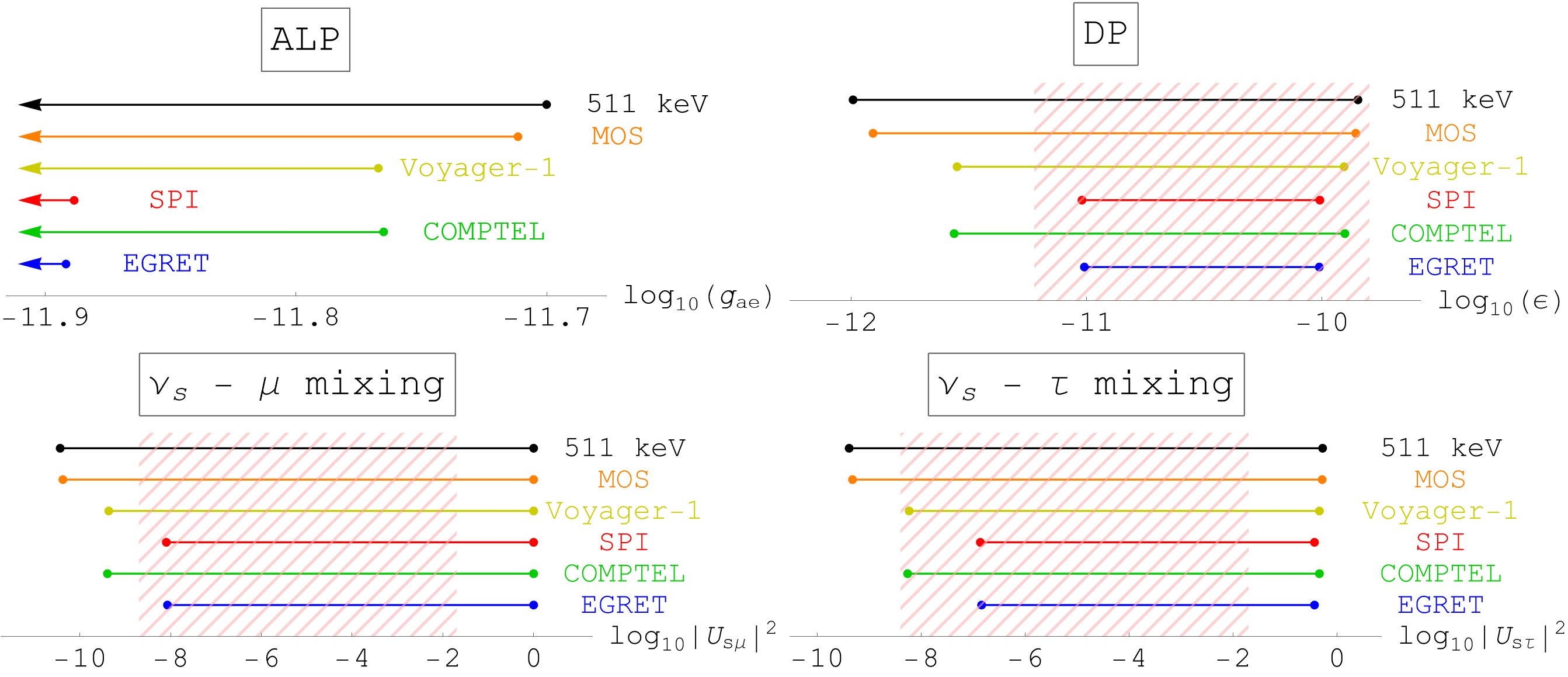}
\end{minipage}
\caption{Limits on the considered FIP models with $m_{X}=10$~MeV at the $95\%$ CL: ALPs (upper left panel), DPs (upper right panel), sterile neutrinos mixed with muon (lower left panel) and tau neutrinos (lower right panel). The most conservative constraints are shown for the SPI bound on the 511~keV line. The color code is in agreement with Fig.~\ref{fig:Limits_Arrow}. Note that the ALP constraint extends to small couplings $g_{ae}\sim\mathcal{O}(10^{-14})$ until the condition in Eq.~\eqref{eq:lowerlimit} is satisfied. The sterile neutrino mixing angle in the lower left panel, in the upper limit, is assumed to saturate to one. The red hatched regions show the parameter space excluded by existing astrophysical constraints~\cite{Sung:2019xie,Calore:2021lih,Carenza:2023old}. Note that there are no current constraints on ALPs and the hatched region for the upper bound on DPs extends to a region of higher coupling not shown in the plot, up to $\epsilon\sim3\times10^{-7}$.}
\label{fig:FIPbound}
\end{figure*}

Here, we briefly mention the possibility of fireball formation due to FIP decay into electron-positron pairs. If this occurs in a relatively small volume, the high density of the plasma might create an environment in which electrons and positrons interact with each other via bremsstrahlung, producing a hot plasma that affects the energy spectrum of the injected electrons and positrons~\cite{Diamond:2023scc,Diamond:2023cto}. In our model-independent analysis, we write the FIP decay length (in the SNe rest frame) as
\begin{equation}
    l_{\rm dec}=l_{0}\frac{E_{X}}{m_{X}}\sqrt{1-\left(\frac{m_{X}}{E_{X}}\right)^{2}}\,,
\end{equation}
where $E_{X}$ is the FIP energy and the decay length $l_{0}$ (in the FIP rest frame) will depend on the FIP mass $m_{X}$ and coupling to electrons. We assume that FIPs are injected with a quasi-thermal spectrum giving rise to the lepton spectrum in Eq.~\eqref{eq:spectrum}, therefore the average FIP energy will be twice $E_{0}$. Moreover, we consider the benchmark values for the other spectral parameters and $m_{X}=10$~MeV. Following Ref.~\cite{Diamond:2023scc}, we are capable of estimating the region of the parameter space, in terms of $N_{e}$ and $l_{0}$ where the fireball formation through bremsstrahlung $ee\to ee\gamma$ happens. 
Applying these results to the considered FIPs, the fireball formation will take place for sterile neutrinos with a mixing $|U_{s\alpha}|^{2}\gtrsim10^{-5}$ and dark photons with $\epsilon\gtrsim10^{-11}$, both cases are already excluded by other SN constraints, in particular for the energy deposition inside low-energy SNe~\cite{Calore:2021lih,Carenza:2023old} (see the red hatched region in Fig.~\ref{fig:FIPbound}).  Furthermore, it is important to remark that the parameter space suitable for the fireball formation is very sensitive to the SN radius. The discussed results are obtained for a nominal radius $3\times10^{12}$~cm, but this parameter varies up to $10^{14}$~cm~\cite{DeRocco:2019njg} and, in this case, the fireball is unlikely to form in the parameter range considered. \\
More precisely, the fireball formation is relevant in the case of SN 1987A, which had a blue supergiant progenitor. The small radius of blue supergiants, around $3\times10^{12}$~cm, would allow a larger fraction of FIPs to decay outside the envelope producing a fireball.
However, the FIP fluxes discussed in this work sum contributions from several SNe that occurred in the past. Thus, the most important contribution to the FIP flux is associated with the much more common red supergiant SN progenitors, with a significantly larger radius about $10^{14}$~cm. The larger radius makes it harder for FIPs to escape the envelope and decay forming a fireball. In this scenario with a red supergiant progenitor for a SN, the fireball does not form at any point in the FIP parameter space. Since an SN originating from a red supergiant progenitor is much more likely than from blue supergiants~\cite{Smartt:2009zr}.
In this first study we neglect the possibility of fireball formation and we defer a more complete analysis, including a fraction of SNe with blue supergiant progenitors, to a future work~\cite{Balajinew}.

\section{Conclusions}
\label{sec:conclusions}

Supernovae (SNe) provide valuable information about the existence and properties of electrophilic Feebly Interacting Particles (FIPs), which are essential for investigating exotic physics. The objective of this research is to analyze and constrain the production of FIPs in SNe using different astrophysical messengers, thereby playing a vital role in exploring the interplay between new physics and astrophysical probes.

Our study focuses on exploring the multimessenger signals generated by FIPs that decay into electron-positron pairs. We aim to demonstrate the effectiveness of this phenomenology in constraining the production and properties of various FIPs, such as axion-like particles, sterile neutrinos, and dark photons. The production of electrons and positrons by FIPs contributes to the overall diffuse background of these particles in our Galaxy. Additionally, when these particles interact with the interstellar gas, they give rise to secondary emissions of photons across a wide spectrum, ranging from X-rays to $\gamma$ rays. By accurately modeling these emissions and considering the $511$~keV signal produced by positron annihilation, we can establish constraints on the total flux of injected electrons and positrons, regardless of the specific FIP model being considered. The discussed phenomenology heavily relies on the description of secondary fluxes. In Ref.~\cite{DelaTorreLuque:2023nhh} we extensively discuss the uncertainties related to the propagation parameters. The main conclusion, that we report here for clarity, is that uncertainties in the source distribution, diffusion coefficient and Galactic halo height, are subleading compared to the uncertainties on the  Alfvèn speed. This is the parameter controlling the reacceleration of charged particles when interacting with Galactic plasma turbulences~\cite{1995ApJ...441..209H, 1998ApJ...509..212S, Cardillo:2019ara, 2022ApJ...941...65B}. The experimental determination of this parameter is challenging and various CR analyses obtain very different results, in the range $V_A\sim0$-$40$~km/s~\cite{delaTorreLuque:2022vhm, Luque:2021nxb, Luque:2021ddh}. As discussed, the Voyager-1 bound is insensitive to this parameter, while the MOS constraint can vary by an order of magnitude in the FIP-electron coupling. For the latter, we remark that the limits presented here (obtained with the benchmark value of $V_A = 13$~km/s) could weaken at most by a factor of $2$, in the most pessimistic case (the non-realistic case of no reacceleration at all), and improve by up to an order of magnitude, in the optimistic case ($V_A = 40$~km/s).
This is the reason why combining different probes allows us to robustly constrain FIP properties, instead of relying on a single observable.

We have significantly improved constraints on FIP models when comparing with previous studies. This includes refining the assessment of uncertainties in the SPI analysis, in order to establish more reliable constraints. Using measurements from Voyager-1, accurately determining how the local electron flux beyond the heliosphere allows us to constrain FIPs, for the first time. Furthermore, we have systematically examined secondary emissions from electron-positron pairs, specifically the inverse Compton and bremsstrahlung processes, by analyzing available datasets from X-ray to $\gamma$-ray energies. These additional data provide stringent limits on electrophilic FIPs. 
In addition, after obtaining limits on the production of these particles and their injection of electrons/positrons, we show how these limits can be translated to concrete parameters defining the properties of axion-like particles, sterile neutrinos, and dark photons. 

\vspace{0.2cm}

In summary, our study emphasizes the importance of galactic signals produced by SNe in the exploration of FIPs. Our journey into the phenomenology of electrophilic FIPs from SN started in Ref.~\cite{DelaTorreLuque:2023nhh}, where we proved the effectiveness of the XMM-Newton data in constraining exotic physics. Moving one step further, in this work we employ a multimessenger approach incorporating observations from diverse cosmic messengers to enhance our comprehension of the FIP phenomenology. The exhaustive analysis that we carried out sets a new standard for the methodologies employed in this type of investigation. Our research establishes a solid groundwork for future inquiries to shed light on the characteristics of FIPs and their significance in astroparticle physics.

\vspace{0.5cm}
\acknowledgements
SB would like to thank Jordan Koechler and Marco Cirelli for useful discussions regarding XMM-Newton data. This article is based upon work from COST Action COSMIC WISPers CA21106, supported by COST (European Cooperation in Science and Technology).
SB is supported by funding from the European Union’s Horizon 2020 research and innovation programme under grant agreement No.~101002846 (ERC CoG ``CosmoChart'') as well as support from the Initiative Physique des Infinis (IPI), a research training program of the Idex SUPER at Sorbonne Universit\'{e}. The work of PC is supported by the European Research Council under Grant No.~742104 and by the Swedish Research Council (VR) under grants  2018-03641 and 2019-02337. 
PDL is supported by the European Research Council under grant 742104 and the Swedish National Space Agency under contract 117/19
This project used computing resources from the Swedish National Infrastructure for Computing (SNIC) under project Nos. 2021/3-42, 2021/6-326, 2021-1-24 and 2022/3-27 partially funded by the Swedish Research Council through grant no. 2018-05973.

\bibliographystyle{bibi}
\bibliography{biblio.bib}
\end{document}